\documentclass[10pt]{article}
\usepackage{amssymb,latexsym}
\usepackage{amsmath,amsbsy}
\usepackage{epsfig}
\usepackage{amscd} 
\usepackage{feynmf}
\DeclareMathOperator{\pperp}{\mathbf{p}^{\perp}}

\DeclareMathOperator{\Pplus}{P^{+}}
\DeclareMathOperator{\pplus}{p^{+}}
\DeclareMathOperator{\qperp}{\mathbf{q}^{\perp}}

\DeclareMathOperator{\kperp}{\mathbf{k}^{\perp}}
\DeclareMathOperator{\kplus}{k^{+}}

\DeclareMathOperator{\ie}{i \epsilon}
\DeclareMathOperator{\Res}{Res}

\DeclareMathOperator{\dw}{D_{\text{W}}}

\DeclareMathOperator{\pp}{p^{\prime}}
\DeclareMathOperator{\GDA}{\Phi(z, \zeta, s)}

\DeclareMathOperator{\kminus}{k^{--}}
\DeclareMathOperator{\pminus}{p^{--}}
\DeclareMathOperator{\qminus}{q^{--}}
\DeclareMathOperator{\sgn}{Sgn}
\DeclareMathOperator{\siim}{Sgn(Im}
\DeclareMathOperator{\qplus}{q^+}
\DeclareMathOperator{\Gt}{\tilde{G}}
\DeclareMathOperator{\pe}{\mathcal{P}}
\DeclareMathOperator{\qu}{\mathcal{Q}}
\DeclareMathOperator{\psiket}{|\psi\rangle}
\DeclareMathOperator{\psitwoket}{|\psi_{2}\rangle}
\DeclareMathOperator{\psiquket}{|\psi_{\qu}\rangle}

\DeclareMathOperator{\psitwobra}{\langle\psi_{2}|}

\begin{document}

\title{\hskip7cm NT@UW 02-0011\\
Light front Bethe-Salpeter equation applied to form factors, generalized parton distributions and generalized distribution amplitudes}
\author{ B.~C.~Tiburzi\footnote{Email: bctiburz@u.washington.edu} \; and G.~A.~Miller\\
        Department of Physics\\  
University of Washington\\      
Box $351560$\\  
Seattle, WA $98195-1560$ }
\date{\today}
\maketitle

\begin{abstract}
We present an intuitive, light front reduction scheme for the Bethe-Salpeter equation that consists of iterations of the (light front) energy
dependent covariant equation followed by an instantaneous approximation. This scheme is equivalent to and motivates the form of a standard reduction
that reproduces light front, time ordered perturbation theory for bound states.
We use this three-dimensional light front formalism to compute electromagnetic form factors between bound states to show how higher, light front Fock components arise from the covariant Bethe-Salpeter wave function.  Moreover, we show that non-wave function vertices, which arise after integration over the minus momenta, are completely removed in favor
of higher Fock states when one correctly takes into account all contributions at a given order in perturbation theory. 
Such non-wave function vertices are shown only to occur in models in which the interactions are instantaneous. These models lack true higher Fock states.
The normalization of the reduced wave function is derived from the covariant framework and related to nonvalence probabilities 
using familiar Fock space projection operators. Using a simple model, we obtain expressions for generalized parton 
distributions that are continuous. The nonvanishing of these distributions at the crossover between kinematic regimes (where the plus component 
of the struck quark's momentum is equal to the plus component of the momentum transfer) is tied to higher Fock components. 
Lastly we apply the light front reduction to timelike form factors and derive expressions for the generalized distribution amplitudes in this model. 
\end{abstract}

\section{Introduction}
More than a half century ago, Dirac's paper on the forms of relativistic dynamics \cite{Dirac:1949cp} introduced the front form
Hamiltonian approach.  Applications to quantum mechanics and field theory were overlooked at the time due to the 
appearance of covariant perturbation theory, in which the special role played by time was supplanted by Lorentz invariance. 
Front form dynamics gradually resurfaced, first under the guise of the infinite momentum frame (in which covariant calculations
could be achieved more easily if the processes were viewed at the speed of light) and later as quantization on the light front plane
(or null plane). 

The reemergence of front form dynamics was largely motivated by simplicity as well as physicality. The light
front approach has the largest stability group \cite{Leutwyler:1977vy} of any Hamiltonian theory, that is light front dynamics
maintains the smallest number of interaction dependent generators of the Poincar\'e algebra. In highly relativistic calculations the utility is clear:
Lorentz boosts on the light front are merely kinematical; whereas in the instant form, boosts depend on the dynamical evolution of the system. 
The physical motivation for hard scattering processes, such as deeply inelastic scattering, was the infinite momentum frame's utility 
in describing partonic subprocesses. Today this connection to light front dynamics is transparent: hard scattering processes probe 
a light cone correlation of the fields. Not surprisingly, then, many perturbative QCD applications can be treated on the light front, see e.g. 
\cite{Lepage:1980fj}. Outside this realm, physics on the light cone has been extensively developed for non-perturbative QCD 
\cite{Brodsky:1997de} as well as applied to nuclear physics \cite{Carbonell:1998rj,Miller:2000kv}.

This paper concerns current matrix elements between bound states of two particles in the light front formalism. We approach the topic,
however, from covariant perturbation theory. By projecting covariant quantities onto the light cone, one has a way of deriving
light front amplitudes from field theory, circumventing the often subtle task of formulating the front form of dynamics \emph{ab initio}.
As demonstrated by the tremendous undertaking of \cite{Ligterink:1994tm}, one can derive light front perturbation theory from covariant perturbation 
theory for scattering states, thereby demonstrating their equivalence---including the delicate issue of renormalization. As to 
the issue of light front bound states, a reduction scheme for the Bethe-Salpeter equation recently appeared \cite{Sales:1999ec} 
that produces a kernel calculated in light front perturbation theory. These authors demonstrate their reduction but they do not focus 
on its origin. We adopt this reduction scheme of the Bethe-Salpeter equation below, however, only after suitable intuitive motivation. 
For the purpose of simplicity, we consider only bound states of two scalar \emph{quarks} interacting via the exchange of a massive
scalar. The intuitive reduction scheme is quite similar to one found in \cite{Brodsky:1984vp}, however, we focus of the Bethe-Salpeter 
vertex function and its energy poles. 

Our main consideration is to extend the reduction to current matrix elements to investigate valence and nonvalence contributions in the light
front reduced, Bethe-Salpeter formalism. Thus we calculate our model's form factor; moreover, we choose not to exploit a special reference 
frame in which we can neglect Z graphs. 
This enables us to completely investigate their contribution (as well as functional form) which has a variety of applications 
from timelike decays and timelike form factors to generalized parton distributions. These Z graph contributions have haunted light front dynamics 
since the nonvalence properties of the bound state are involved and thus valence wave function models cannot be utilized directly.  
A variety of solutions to this dilemma have been pursued. In \cite{Hwang:2001wu}, the nonvalence piece was treated as a new sort of wave function 
with an extra parameter determined by fitting to data. The approach allows for some phenomenological consistency, although the underlying field 
theoretic nature of the problem is ignored, e.g. Lorentz invariance is not maintained. The authors 
\cite{Ji:2000fy} propose that the light front Bethe-Salpeter equation can be iterated to relate the nonvalence contribution to a valence wave 
function. In lieu of knowing the interaction which performs this crossing, they propose that the nonvalence vertex can be replaced by a constant. 
The reader is then riddled by their insistence that the constant vertex is consistent with data and therefore generally true, and later
falsely assured that the vertex can uniquely be determined by Lorentz invariance (and hence depends only on one variable $\zeta$, perhaps two $\zeta, t$). 
The constant approximation and Lorentz invariance fitting are far from unique since the complete functional dependence (on $\zeta, t, x, \kperp$) 
of the vertex is ignored (and hence true nonvalence correlations are wiped out). The procedure also results in conspicuously discontinuous
distributions which one can trace directly to the constant vertex approximation.  One could just as well choose a vertex with 
some arbitrary dependence on the neglected variables that forces continuity and includes an overall function to be determined from Lorentz invariance. 
This approach is a hybrid of \cite{Hwang:2001wu} and \cite{Ji:2000fy} that has not been pursued, until very recently \cite{Choi:2002ic}.

We, too, tried a similar approach of iterating the reduced Bethe-Salpeter equation \cite{Tiburzi:2001ta} in an attempt to 
rigorously generalize the results of \cite{Einhorn:1976uz}. We used the same simple Lagrangian model used below to 
express the crossed interactions and it seemed one could avoid explicitly using higher Fock space components. Our results, however, 
were not rigorously obtained in perturbation theory. We show below that contributions from non-wave function vertices are exactly canceled  
when one works completely to a given order in perturbation theory. Thus one cannot side step the issue of higher Fock components. 

We begin in section \ref{intuitive} by considering an intuitive reduction scheme for the Bethe-Salpeter equation. The method consists
of a series of iterations of the covariant equation, followed by an instantaneous approximation for the vertex function, which allows for 
all integrals over minus momenta to be performed. The focus concerns the light front energy poles of the Bethe-Salpeter vertex, 
and further iteration before an instantaneous reduction allows for a better approximation to the poles.  
We use this scheme to derive the bound state equation to second order in the interaction, 
making the connection to light front, time ordered perturbation theory. 
We then use this intuitive scheme in section \ref{formal} to motivate the formal reduction carried out by \cite{Sales:1999ec}. Once we have 
the formal scheme, we reverse the logic and  demonstrate how the intuitive scheme arises. Discussion
of the light front wave function's normalization is delayed until Appendix A. 

The formal reduction is then used
to obtain current matrix elements between bound states in section \ref{current}. We first formally construct the gauge invariant current
and then apply it to form factors in perturbation theory. At leading order we are confronted with non-wave function vertices, but in order to appeal
to crossing symmetry we must go to higher order. Considering then the complete next to leading order expressions, we show the non-wave function 
vertex contributions are removed and replaced by higher Fock components.  
In Appendix B, we describe how crossing symmetry can be exploited in the case of an instantaneous interaction, 
such as QCD${}_{1+1}$ \cite{Einhorn:1976uz}.

The immediate application of our development for form factors (in a frame where the plus momentum transfer is non-zero) is to compute generalized parton 
distributions.  We carry this out in section \ref{gpd} where continuous behavior at the crossover is demonstrated. We also make explicit the relation to 
Fock component overlaps in this model.
Connecting back to the intuitive scheme, we notice that the poles of the covariant vertex function generate the higher Fock
components present in our expressions for form factors---and quite naturally the Bethe-Salpeter wave function retains all Fock components (in principle).
This necessarily implies the same is certainly \emph{not} true of the reduced light front wave function. Its lack of light front energy dependence 
restricts it to the valence sector alone. In Appendix C, we derive the higher Fock components for the model directly by using old fashioned, time ordered
perturbation theory for the light front Hamiltonian. Another application is pursued in section \ref{gda}, where we derive expressions for the generalized 
distribution amplitude for this model, which is related to the timelike form factor via a sum rule. 
Such an application is interesting since no Fock space expansion in terms of two-body bound states is possible. Lastly we conclude briefly in 
section \ref{conclusion}.

\section{Intuitive reduction}\label{intuitive}

\begin{figure}
\begin{center}
\epsfig{file=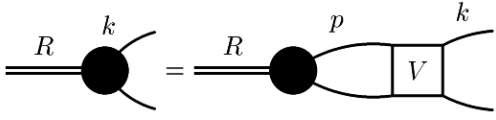}
\end{center}
\caption{Diagrammatic representation of the Bethe-Salpeter equation. The blob represents the vertex function $\Gamma$.}
\label{fBS}
\end{figure}

We start by writing down the covariant equation for the meson vertex function $\Gamma$. It satisfies a simple 
Bethe-Salpeter equation \cite{Schwinger:1951ex}, first schematically as operators (see Figure \ref{fBS}):
\begin{equation} \label{Gamma}
\Gamma = V G \Gamma,
\end{equation}
where $G$ is the full, renormalized, two-particle disconnected Green's function (which can be expressed as a product of two, renormalized, single-particle propagators). Here $V$ is the \emph{interaction potential}, which is the irreducible two-to-two scattering kernel. Let us denote the meson four-momentum by $R$, the quark's by $k$, and the physical masses by $R^2 \equiv M^2, k^2 \equiv m^2$. Since we take our quarks to be scalar particles, the renormalized, single particle propagator has a Klein-Gordon form:
\begin{equation}
d(k) = \frac{i}{(k^2 - m^2)[1+ (k^2 - m^2)f(k^2)]+ i \epsilon},
\end{equation}
where the residue is $i$ at the physical mass pole and the function $f(k^2)$ characterizes the renormalized, one-particle irreducible self interactions. 
We shall assume there are no poles (besides at the physical mass) in the propagator. This is consistent with model studies \cite{Roberts:1994dr}. 
Since our analysis hinges on poles, we can neglect $f(k^2)$ in what follows by noting that to add it back in, we merely evaluate at the relevant poles (since all integrations in question will be calculated from residues). 

In the momentum representation, the Bethe-Salpeter equation then appears as 
\begin{equation} \label{BS}
\Gamma(k,R) =  i \int \frac{d^4p}{(2\pi)^4} V(k,p) \Psi_{BS}(p,R),
\end{equation}
in which we have defined the Bethe-Salpeter wave function $\Psi_{BS}$ as
\begin{equation}\label{psiBS}
\Psi_{BS}(k,R) = G(k,R) \Gamma(k,R),
\end{equation}
with $G(k,R) = d(k) d(R-k)$. Since we will ultimately be concerned with soft physics probed in hard processes, we imagine the initial conditions of our system are specified on the hypersurface $x^+= 0$.  
(We define the plus and minus components of any vector by $x^\pm = \frac{1}{\sqrt{2}}(x^0 \pm x^3)$.) Correlations between field operators 
evaluated at equal light cone time turn up in hard processes and thus this choice of initial surface is natural. In order to project out
the initial conditions (wave functions, \emph{etc}.) of our system, we must perform the integration over the Fourier conjugate to $x^+$, namely $k^-$.
For instance, our concern is with the light front wave function defined as the projection of the covariant Bethe-Salpeter wave function onto $x^+ = 0$,
\begin{equation}
\psi(\kplus, \kperp, R) = 2 R^+ x(1-x) \int \frac{dk^-}{2\pi} \Psi_{BS}(k,R).
\end{equation}

Looking at Eq. \eqref{psiBS}, in order to project the wave function exactly, we must know the analytic structure of the bound state vertex function. If the vertex function $\Gamma(k,R)$ had no poles in $\kminus$, then our task would be simple: the light front projection of $\Psi_{BS}$ would pick up contributions only from the poles of the propagator $G(k,R)$. Next we observe from Eq. \eqref{BS}, that the $\kminus$ dependence of the interaction $V(k,p)$ must give rise to the $\kminus$ poles of the vertex function $\Gamma(k,R)$. 

If we imagine an interaction which is dominated by a piece instantaneous in light front time $x^+$ (i.e.~independent of the light front energy $\kminus$), the vertex function will similarly be independent light front energy. Under these assumptions, the projection onto the initial surface can be carried out yielding a zeroth order approximation to the light front wave function. (The equal time version of this approximation was first carried out by Salpeter \cite{Salpeter:1952ib} but by specifically using the instantaneous Coulomb interaction.) Before worrying about how to correct this approximation, let us spell out the procedure and form of the zeroth order wave function. 

The vertex function $\Gamma$ is replaced by its instantaneous approximation $\gamma$ which must be three-dimensional. Thus $\gamma(k,R)$ depends upon 
$k^2 = 2 \kplus \kminus_{\text{on}} - \kperp^2 = m^2$, $R^2 = M^2$, and $k \cdot R$, where
\begin{align}
k \cdot R & = \kminus_{\text{on}} R^{+} + \kplus R^{-} - \kperp \cdot \mathbf{R}^{\perp} \notag \\
	  & = \frac{1}{2 x} (m^2 + x M^2  + \kperp_{\text{rel}}^2), \notag
\end{align}
with $x = \kplus /R^{+}$ and $\kperp_{\text{rel}} = \kperp - x \mathbf{R}^{\perp}$. We shall write this dependence as $\gamma = \gamma(x,\kperp_{\text{rel}}|M^2)$ and often for simplicity omit the relative subscript on the transverse momentum. The projection is then carried out simply as noted by integrating the propagator over $\kminus$, 
\begin{equation}
 \int d\kminus d(k) d(R-k) = \int \frac{d\kminus}{2\kplus 2(\kplus - R^{+})} \frac{i}{\kminus - \kminus_{a}} \frac{i}{\kminus - \kminus_{b}}
\end{equation}
where
 \begin{equation} \label{valpoles}
 \begin{cases}
 \kminus_{a} = \kminus_{\text{on}} - \frac{\ie}{2 \kplus}\\
 \kminus_{b} = R^{-} + (k - R)^{-}_{\text{on}} - \frac{\ie}{2 (\kplus - R^{+})}.
 \end{cases} 
 \end{equation} 
 The above integral is performed using the Cauchy integral theorem. For $\kplus <0$ both poles $a$ and $b$ lie in the upper-half complex $\kminus$ plane while for $\kplus > R^{+}$ both lie in the lower-half plane. In both cases, we can close the contour to avoid enclosing any poles and the integral vanishes. On the other hand, for $0< \kplus < R^{+}$ closing the contour in the upper-half plane picks up
 \begin{align} \label{g}
 \int d\kminus d(k) d(R-k) & = 2 \pi i \Res(\kminus_{b}) \theta[x(1-x)] \notag \\ 
 & = \frac{\pi i }{x(1-x) R^+} \dw (x,\kperp_{\text{rel}}|M^2) \theta[x(1-x)],
 \end{align}
where we have defined the Weinberg propagator as
\begin{equation}
\dw (x, \kperp_{\text{rel}}|M^2) = \frac{1}{2 R^+(\kminus_{b} - \kminus_{a})}  = \frac{1}{M^2 - \frac{\kperp_{\text{rel}}^2 + m^2}{x(1-x)}}.
\end{equation}
 The zeroth order, light front wave function is then
 \begin{equation}
 \psi^{(0)} = i \dw(x,\kperp_{\text{rel}}|M^2) \gamma(x,\kperp_{\text{rel}}|M^2) \theta[x(1-x)]
 \end{equation}
and hence is a function of only the relative momenta.

 To better approximate the wave function, we need to correct for using the instantaneous approximation. We can achieve this systematically by iterating the 
 covariant Bethe-Salpeter equation for $\Gamma$. A better approximation to $\psi$
 would be iterating once and then replacing $\Gamma$ with the instantaneous vertex $\gamma$. By retaining energy dependence in $V$, one is allowing for minus momentum poles in $\Gamma$, though we still must make an instantaneous approximation in order to carry out the integration. The next better approximation uses two iterations before using an instantaneous vertex replacement. Schematically the chain of successively better light front wave functions can be depicted as
 \begin{align}
 \psi^{(0)} & = \int G \gamma \notag \\
 \psi^{(1)} & = \int G V G \gamma \notag \\
 \psi^{(2)} & = \int (G V)^2 G \gamma \notag \\
 \begin{CD}
 @VVV
 \end{CD} \notag \\
 \psi^{(j)} & = \int  (G V)^j G \gamma. \notag 
 \end{align}

 Of course the exact light front wave function is $\psi^{(\infty)}$ for which the poles of the vertex remain unaltered. At any level in the scheme, $\psi^{(j)}$ will depend upon the $j-1$ previous approximations. To derive a bound state integral equation for $\psi^{(j)}$, one merely replaces all $\psi^{(k)}$, $k<j$ with $\psi^{(j)}$. This is possible since in the equation for $\psi^{(j)}$, a term with $\psi^{(k)}$ will be multiplied by $j-k$ powers of $V$. Imagining that the interaction is weakly retarded, we can upgrade $\psi^{(k)}$ to $\psi^{(j)}$ for free, since the difference introduces terms of order at least $j+1$ in the interaction which are irrelevant: $\psi^{(j)}$ is valid only to order $j$.\footnote{In this way we are making a perturbative expansion for the
reduced kernel and not counting powers of the interaction (which must be large $GV \sim 1$ to produce a bound state). 
This will be made clear for the ladder approximation to the kernel below.} 

 To illustrate this intuitive scheme, we desperately require an example. Indeed to say anything less than general, we must know the $\kminus$ dependence of the interaction. We therefore adopt a weakly coupled, one-boson exchange model for $V$ (the so-called ladder approximation). Supposing the boson mass is $\mu$ and the coupling constant $g$, we have
 \begin{align} \label{ladder}
 V(k,p) & =  \frac{- g^2}{(p-k)^2 - \mu^2 + \ie}  \notag \\
 	& = \frac{- g^2}{2(\pplus - \kplus)}\frac{1}{\pminus - \pminus_{v}},
 \end{align}
 where the energy pole (with respect to $p$) of the interaction is 
 \begin{equation} \label{pv}
 \pminus_{v} = \kminus + \frac{(\pperp -\kperp)^{2}+\mu^2}{2 (\pplus - \kplus)} -\frac{\ie}{2(\pplus - \kplus)}.
 \end{equation}
 With this model, we will work out two non-trivial orders of the reduction scheme.  

\subsection{First order interaction}
\begin{figure}
\begin{center}
\epsfig{file=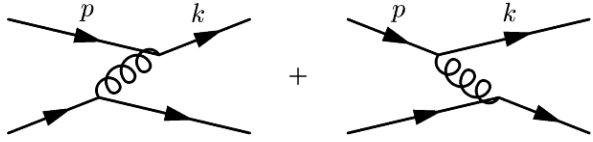}
\end{center}
\caption{Diagrammatic representation of the one-boson exchange potential $V$ appearing in Eq. \eqref{OBE}.}
\label{fOBEP}
\end{figure}

 To first order in a weakly retarded interaction, we have
\begin{equation} \label{firsto}
\psi^{(1)}(x,\kperp) = 2 R^+ x(1-x) \int \frac{d\kminus}{2 \pi} G(k,R) i \int \frac{d^4p}{(2\pi)^4}  V(k,p) G(p,R) \gamma(y,\pperp_{\text{rel}}|M^2),
\end{equation}
where we have defined $y = \pplus/R^+$. Having inserted the instantaneous vertex $\gamma$ after one iteration of the Bethe-Salpeter equation, the integral
over $\pminus$ can now be performed in addition to the $\kminus$ integral. To begin, first consider the $\pminus$ integration. The integrand has three poles
(two from the propagator and one from the interaction)
\begin{equation} \label{polesv1}
\begin{cases}
\pminus_{a} = \pminus_{\text{on}} - \frac{\ie}{2 R^+}\frac{1}{y} \\
\pminus_{b} = R^- + (p - R)^{-}_{\text{on}} - \frac{\ie}{2 R^+} \frac{1}{y-1} \\
\pminus_{v} = \kminus + \frac{(\pperp -\kperp)^{2}+\mu^2}{2 (\pplus - \kplus)} -\frac{\ie}{2 R^+}\frac{1}{y-x},
\end{cases}
\end{equation}  
which have imaginary parts that switch sign depending on the size of $x$ and $y$. In Table \ref{t:v1}, we enumerate the possible combination of signs 
of the poles' imaginary parts. 
\begin{table}
\begin{center}
	\begin{tabular}{|| l || c | c | c | c | c | c ||}
	\hline
	Case     &   y$<$0   &   y$<$0   & 0$<$y$<$1   &  0$<$y$<$1   &  y$>$1   & y$>$1  \\
                 &   y$<$x   &   y$>$x   &  y$<$x    &  y$>$x     &  y$<$x   & y$>$x  \\ 
	\hline
	Positive &    $a$ $b$ $v$ &  $a$ $b$    &  $b$ $v$    &  $b$       &  $v$     &      \\ 
 	\hline  
	Negative &         &   $v$     & $a$       &  $a$ $v$     &  $a$ $b$   & $a$ $b$ $v$ \\
	\hline	
	\end{tabular}
\end{center}
\caption{All possible combinations of signs of the poles' imaginary parts for the first order, light front wave function. The $\pminus$ poles 
are labeled as they appear in Eq. \eqref{polesv1}.}  
\label{t:v1}
\end{table} 
So for example, when $y<0$ and $y<x$, we close the contour in the lower-half plane and the integral vanishes, as is the case additionally 
when $y>1$ and $y>x$. To simplify our work, we can assume that contributions to $\psi$ will require $0<x<1$.\footnote{If we do not assume this, 
the additional residues with respect to $\pminus$ which we pick up in Table \ref{t:v1} turn out to evaluate to zero when integrated with respect to $\kminus$. In this way, we 
need not assume the wave function $\psi^{(j)}$ is zero outside $0<x<1$; extra computation will demonstrate that it indeed vanishes.} Thus two regions
are eliminated due to impossibilities: $(y<0) \cap  (y>x) = \phi $ and $ (y >1) \cap (y < x) = \phi$. The integral is 
\begin{multline} \label{preint}
2 \pi i \theta[y(1-y)] \Big( \theta(y - x) \Res(\pminus_{b}) - \theta(x - y) \Res(\pminus_{a}) \Big)  \\
= 2 R^+ x(1-x) \int \frac{d\kminus}{2\pi} G(k,R) \int \frac{dy d\pperp}{2 (2 \pi)^3 y(1-y)} 
\tilde{V}(\kminus,x,\kperp;y,\pperp) (-i) \psi^{(0)}(y,\pperp_{\text{rel}}),
\end{multline}
where we used $\theta[y(1-y)] \dw(y,\pperp|M^2) \gamma(y,\pperp|M^2) = - i \psi^{(0)}(y,\pperp |M^2)$ and defined
\begin{equation}
\tilde{V}(\kminus,x,\kperp;y,\pperp) = \frac{g^2}{2 R^+ (y -x)} \Bigg( \frac{\theta(y -x)}{\pminus_{b} - \pminus_{v}} 
+ \frac{\theta(x-y)}{\pminus_{a} - \pminus_{v}} \Bigg). 
\end{equation}
Above we commented that $\psi^{(1)}$ would depend upon the lower order approximation $\psi^{(0)}$ and this is indeed the case. To derive a bound state 
equation for the first order, light front wave function, we merely realize that changing $\psi^{(0)}$ to $\psi^{(1)}$ above 
results in corrections of second order in the retarded interaction. To derive this three-dimensional bound state equation, we need to perform 
the $\kminus$ integral. $\tilde{V}$ retains $\kminus$ dependence through $\pminus_{v}$ and hence a renaming scheme is necessitated.
Let us define
\begin{equation} \label{vpoles}
\begin{cases}
\kminus_{v_{a}} = \frac{(\kperp - \pperp)^2 + \mu^2}{2R^+(x - y)} + \pminus_{\text{on}} + \frac{\ie}{2 R^+}\frac{x}{(y-x)y}  \\
\kminus_{v_{b}} = \frac{(\kperp - \pperp)^2 + \mu^2}{2R^+(x - y)} + R^- + (p - R)^-_{\text{on}} + \frac{\ie}{2R^+}\frac{x-1}{(y-x)(y-1)}, 
\end{cases}
\end{equation}
which renders
\begin{equation} \label{vtilde}
\tilde{V}(\kminus,x,\kperp;y,\pperp) = \frac{g^2}{2 R^+(x-y)} \Bigg( \frac{\theta(y-x)}{\kminus - \kminus_{v_{b}}} + 
\frac{\theta(x-y)}{\kminus - \kminus_{v_{a}}}  \Bigg).
\end{equation}
Four poles now confront us in performing the second minus momentum integration: $\{\kminus_{a}, \kminus_{b}, \kminus_{v_{a}}, \kminus_{v_{b}}\}$.
All possible combinations of signs of their imaginary parts are listed in Table \ref{t:v11}. 
\begin{table}
\begin{center}
	\begin{tabular}{|| l || c | c | c | c | c | c ||}
	\hline
	Case     &   x$<$0   &   x$<$0   & 0$<$x$<$1   &  0$<$x$<$1   &  x$>$1   & x$>$1  \\
                 &   y$<$x   &   y$>$x   &  y$<$x      &  y$>$x       &  y$<$x   & y$>$x  \\ 
	\hline
	Positive &   $a$ $b$ $v_{a}$ &  $a$ $b$ $v_{b}$    &  $b$     &  $b$ $v_{b}$       &       &      \\ 
 	\hline  
	Negative &         &        & $a$ $v_{a}$    &  $a$      &  $a$ $b$ $v_{a}$  & $a$ $b$ $v_{b}$ \\
	\hline	
	\end{tabular}
\end{center}
\caption{All possible combinations of signs of the poles' imaginary parts for the first order bound state equation. The $\kminus$ poles are labeled as they 
appear in Eqs. \eqref{valpoles} and \eqref{vpoles}.}  
\label{t:v11}
\end{table} 
Using Cauchy, we end up only with two residues to evaluate yielding the desired bound state equation
\begin{equation} \label{wavefunction}
\psi^{(1)}(x,\kperp_{\text{rel}}) = \dw(x,\kperp_{\text{rel}}|M^2) \int \frac{dy d\pperp_{\text{rel}}}{2(2\pi)^2 y(1-y)} V(x,\kperp_{\text{rel}};y,\pperp_{\text{rel}}) \psi^{(1)}(y,\pperp_{\text{rel}}),
\end{equation}
for the first order interaction $V$ defined by
\begin{equation} \label{OBE}
V(x,\kperp_{\text{rel}};y,\pperp_{\text{rel}}) = \theta(x-y) \tilde{V}(\kminus_{b},x,\kperp;y,\pperp)  
+ \theta(y-x) \tilde{V}(\kminus_{a},x,\kperp;y,\pperp).
\end{equation} 
The equation takes the form of the Weinberg equation \cite{Weinberg:1966jm} for the light front time ordered, one-boson exchange potential, since, for example
\begin{equation}
\frac{ g^2 \theta(y-x)}{2 R^+ (y - x)} \tilde{V}(\kminus_{a},x,\kperp;y,\pperp) ^{-1} = R^- - \kminus_{\text{on}} - \frac{(\pperp-\kperp)^2 + \mu^2}{2 R^+ (y - x)} - (R - p)^{-}_{\text{on}}
\end{equation}
is the energy denominator corresponding to the intermediate quark with momentum $p$ emitting a boson which is later absorbed by the final state quark 
(of momentum $k$). Both light front time ordered contributions to $V$ in Eq. \ref{OBE} are depicted in Figure \ref{fOBEP}.
Therefore, at leading order in the retarded interaction, we have shown the first step in the iterative scheme produces the correct potential computed
from time ordered perturbation theory.  

\subsection{Second order interaction}
We now illustrate the procedure at second order. At two iterations, we have the following equation for the approximate light front wave function
\begin{equation} \label{so}
\psi^{(2)} = 2 R^+ x(1-x) i^2 \int \frac{d\kminus}{2\pi} \; \frac{d^4p}{(2\pi)^4} \; \frac{d^4q}{(2\pi)^4} G(k,R)  V(k,p) G(p,R) V(p,q) G(q,R) \gamma(z,\qperp_{\text{rel}}|M^2),
\end{equation}
where $z \equiv \qplus/R^+$. At this point, there are a few different ways to proceed. We could first perform the $\qminus$ integral using our previous result Eq. \eqref{vtilde} (with the necessary renaming of variables) and then the $\kminus$ integral. Carrying out the integration in this particular order shows one that contributions to $\psi^{(2)}$ are from the region where $0<z<1$. The resulting integrand is quite complicated and further integration and interpretation in terms of time ordered graphs require 
algebraic gymnastics. Furthermore, rather delicate cancellations are required to show that the regions $y>1$ and $y<0$ make a vanishing contribution. We shall not take this route, however, since the integrals all converge, each route must yield the same answer. Hence we shall use the fact that only $0<z<1$ contributes to perform instead the $\pminus$ integral first. There will be fewer combinations to enumerate and the $\pminus$ integral is essentially an example considered by Ligterink and Bakker \cite{Ligterink:1994tm}, who prove the equivalence of covariant perturbation theory and light-front time-ordered perturbation theory for scattering states.   The difference is that we do not assume that contributions to the bound state wave function come from plus momentum fractions between zero and one, rather we derive this fact.

There are four poles relevant to the $\pminus$ integral: two from the propagator ($\pminus_{a}$ and $\pminus_{b}$) and the remaining from the two iterations of the interaction ($\pminus_{v}(\kminus)$ and $\pminus_{v}(\qminus)$, see Eq. \eqref{pv} and make the obvious replacement of variables). For ease we shall abbreviate the latter as $\pminus_{v_{k}}$ and $\pminus_{v_{q}}$, respectively.
The signs of the imaginary parts of the poles are: $\siim(\pminus_{a})) = - \sgn y$, $\siim(\pminus_{b})) = - \sgn (y - 1)$, $\siim(\pminus_{v_{k}})) = - \sgn (y - x)$ and $\siim(\pminus_{v_{q}})) = - \sgn (y - z)$. Given these and $0<z<1$, Table \ref{t:v2} lists all possible combinations of signs depending on $x,y$ and $z$. 
\begin{table}
\begin{center}
	\begin{tabular}{|| l || l || c | c | c | c ||} \hline
        & $\siim(\text{pole}))$ &  $y>x$  & $y>x$ & $y<x$ & $y<x$ \\
	  &  &  $y>z$  & $y<z$ & $y>z$ & $y<z$ \\ \hline
   $y<0$& Positive & $\phi$ & $a$ $b$ $v_{q}$ & $\phi$ & $a$ $b$ $v_{k}$ $v_{q}$ \\ \hline
	& Negative & $\phi$       &  $v_{k}$ &  $\phi$      &    \\ \hline \hline
   $0<y<1$ & Positive & $b$ & $b$ $v_{q}$ & $b$ $v_{k}$ &  $b$ $v_{k}$ $v_{q}$ \\ \hline
	     & Negative & $a$ $v_{k}$ $v_{q}$ &  $a$ $v_{k}$ & $a$ $v_{q}$ & $a$ \\ \hline \hline
   $y>1$	& Positive & & $\phi$ & $v_{k}$ & $\phi$ \\ \hline
            & Negative & $a$ $b$ $v_{k}$ $v_{q}$ &  $\phi$      & $a$ $b$ $v_{q}$ & $\phi$
\\  \hline  
	\end{tabular}
\end{center}
\caption{All possible combinations of signs of the poles' imaginary parts for the second order bound state equation using $0<z<1$. The $\pminus$ poles are labeled as they appear below Eq. \eqref{so}. A $\phi$ denotes an empty intersection of the restricted regions of $x,y$ and $z$.}  
\label{t:v2}
\end{table} 
We anticipate that the contributions from $y$ outside of $(0,1)$ will vanish. Let us then tackle these first. 

For $y<0$, only one region potentially contributes to the integral: $y>x$ and $y<z$. Since $0<z<1$, the last inequality is automatically satisfied. 
The first two inequalities, however, constrain $x<0$. Closing the contour in the lower-half plane, we pick up only the residue at $\pminus_{v_{k}} = 
\pminus_{v}(\kminus)$ which spreads the $\kminus$ dependence to more denominators. There are now five poles in $\kminus$: $\kminus_{a},\kminus_{b}$ 
which come from the propagator $G(k,R)$, two poles which come from evaluating the propagator $G(p,R)$ at $\pminus_{v_{k}}$, 
and a final pole from evaluating $V(p,q)$ at $\pminus_{v_{k}}$. The exact form of these poles is of no concern, only their imaginary parts. 
The poles of the propagator $G(k,R)$ have imaginary parts $- \frac{\ie}{x}$ and $ - \frac{\ie}{x -1}$ which are both positive since $x$ must be negative.
The remaining poles' imaginary parts are shifted by evaluating at $\pminus_{v_{k}}$ since it has an imaginary part $-\frac{\ie}{y-x}$. Thus the poles which come from $G(p,R)$ have imaginary parts $-\frac{\ie}{y} + \frac{\ie}{y-x}$ and $-\frac{\ie}{y-1} + \frac{\ie}{y-x}$ which are again both positive since
$y<0$ and $y>x$. Finally the interaction $V(p,q)$ evaluated at $\pminus = \pminus_{v_{k}}$ has a $\kminus$ pole with imaginary part 
$-\frac{\ie}{y-z} + \frac{\ie}{y-x}$ and since $y<z$ and $y>x$, this is positive. When integrating over $\kminus$, all poles have positive imaginary parts
and thus the integral vanishes in this region.

On the other hand, when $y>1$ only one region can potentially make a contribution: $y<x$ and $y>z$. Quite similarly to the above, since $0<z<1$ we must have
$x>1$. Closing the contour in the upper-half plane we pick up the same residue, namely at $\pminus_{v_{k}}$. Then in proceeding to evaluate the 
$\kminus$ integral, we have the same poles as before. This time however all the relevant signs are reversed and consequently all $\kminus$ poles lie in the lower-half plane.
The contribution from this region is also zero and hence the only possible contributions to the second order wave function are truly from $0<y<1$.

\begin{figure}
\begin{center}
\epsfig{file=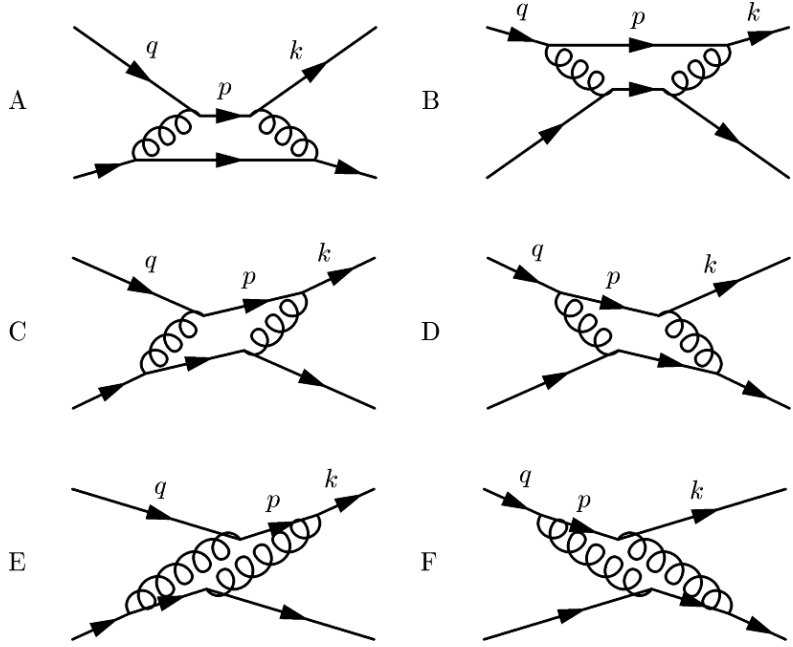}
\end{center}
\caption{Light front time ordered diagrams contributing to the second order wave function. Notice that diagrams A-D can be summed into the square of the 
first order interaction.}
\label{fv2}
\end{figure}

Referring back to Table \ref{t:v2}, there are four distinct regions that will make a contribution to the wave function for $0<y<1$. Two are considerably simpler than the others since we can close the contour to enclose only a single pole. We shall proceed with these. When $y>x$ and $y>z$, the pole in the upper-half plane gives us $+2 \pi i \Res(\pminus_{b})$. Since this pole did not stem from either interaction, there is no spreading of $\kminus$ or $\qminus$ dependence to other denominators. As a result, the two minus momentum integrals factorize into separate integrals both involving a propagator and a single interaction. Ignoring the overall factors, the relevant terms are
\begin{equation} \label{inter}
\frac{1}{(\kminus - \kminus_{a})(\kminus - \kminus_{b})(\kminus - \kminus_{v_{b}})}\; \frac{1}{\pminus_{b} - \pminus_{a}} \; 
\frac{1}{(\qminus - \pminus_{b})(\qminus - \qminus_{a})(\qminus - \qminus_{b})}.
\end{equation}  
Notice the factor of $\dw(y,\pperp|M^2)$ in the middle corresponding to free particle propagation between interactions. The $\kminus$ integral is now identical to that in Eq. \eqref{preint} provided we include the factor $\theta(y - x)$. Thus $0<x<1$ and we pick up a contribution from the pole $\kminus_{a}$. The $\qminus$ integral is essentially the same picking up only the residue at $\qminus_{a}$. Taking these residues of Eq. \eqref{inter} gives us the time ordered diagram A of Figure \ref{fv2}. Parallel to the case just considered, the region $y<x$ and $y<z$ exhibits the same features. We pick up the residue of a single pole, $\pminus_{a}$, which does not contaminate the other terms with $\kminus$ or $\qminus$ dependence. The minus momentum integrals again factorize with a Weinberg propagator sitting between the two interactions.
The light front time ordered graph B in Figure \ref{fv2} corresponds to this contribution. 

The last two regions of integration are more complicated for two reasons. Firstly we pick up the residues of two poles, one of which spreads minus momentum dependence to other energy denominators. Secondly, each residue alone does not correspond to a time ordered graph, rather there are two graphs entangled in their sum. Consider the case where $y>x$ and $y<z$ for which we close the contour in the upper-half plane to pick up (considering only the pieces of the integrand altered by evaluating at the $\pminus$ poles)
\begin{align} 
\Res(\pminus_{b}) + \Res(\pminus_{v_{q}}) & \propto \frac{1}{(\pminus_{b} - \pminus_{v_{k}})(\pminus_{b} - \pminus_{a})(\pminus_{b} - \pminus_{v_{q}})} + \frac{1}{(\pminus_{v_{q}} - \pminus_{a})(\pminus_{v_{q}} - \pminus_{b})(\pminus_{v_{q}} - \pminus_{v_{k}})} \notag \\
& = - \frac{1}{(\pminus_{b} - \pminus_{v_{k}})(\pminus_{b} - \pminus_{a})(\pminus_{v_{q}} - \pminus_{v_{a}})} - \frac{1}{(\pminus_{b} - \pminus_{v_{k}})(\pminus_{v_{q}} - \pminus_{a})(\pminus_{v_{q}} - \pminus_{v_{k}})}. \label{alg} 
\end{align}

After the algebraic manipulation that brings us to the second line, the first term has been cast in a familiar form featuring the free propagator between interactions. The $\kminus$ and $\qminus$ integrals then factorize into forms considered previously. Evaluating these integrals yields an expression corresponding to the time ordered diagram C in the Figure. The second term, however, has one energy denominator $(\pminus_{v_{q}} - \pminus_{v{k}})^{-1}$ which has both $\kminus$ and $\qminus$ dependence brought about by two iterations of the interaction. 

Of the two remaining integrals, consider first the one over $\qminus$. The propagator $G(q,R)$ sitting in Eq. \eqref{so} brings along poles $\qminus_{a}$ and $\qminus_{b}$. The first has an 
imaginary part $-\frac{\ie}{z}$ which is negative for $0<z<1$, while the second has a positive imaginary part $-\frac{\ie}{z-1}$. The other two poles are found in Eq. \eqref{alg}. One is of a familiar form since $\pminus_{v_{q}} - \pminus_{a} = \qminus - \qminus_{v_{a}}$ where $\siim(\qminus_{v_{a}})) = -\frac{\ie \; z}{y(z-y)}$. This imaginary part has a negative sign since $0<z<1$, $0<y<1$ and $y<z$. The last $\qminus$ pole which stems from the denominator $\pminus_{v_{q}} - \pminus_{v_{k}}$ has an imaginary part $\frac{\ie}{x-y} + \frac{\ie}{y-z}$ which is also negative. Hence we pick up $+2\pi i \Res(\qminus_{b})$ by closing the contour in the upper-half plane. Collecting all the terms we have organizes us to perform the last integration over $\kminus$
\begin{equation}
\frac{ \theta(y-x) \theta[y(1-y)] \theta[z(1-z)] \theta(z-y) \tilde{V}(\qminus_{b},z,\qperp;y,\pperp) \dw(z,\qperp|M^2)}{(\kminus - \kminus_{a} )(\kminus - \kminus_{b})(\pminus_{v}(\kminus) - \qminus_{v_{b}})(\kminus - \kminus_{v_{b}})}.
\end{equation}
Now since $y>x$, we must have $x<1$. When $x<0$ all four $\kminus$ poles have positive imaginary parts and when $0<x<1$ only $\kminus_{a}$ has a negative imaginary part. Hence finishing up the integral requires $-2\pi i \Res(\kminus_{a})$ and we have additionally shown $0<x<1$. The result takes the form (with $\theta[x(1-x)] \theta[y(1-y)] \theta[z(1-z)]$ understood)
\begin{equation} \label{mess}
 \dw(x, \kperp|M^2) \theta(y-x) \tilde{V}(\kminus_{a}, x, \kperp; y,\pperp) 
V_{\text{F}}(x,\kperp;y,\pperp;z,\qperp) 
\theta(z-y) \tilde{V}(\qminus_{b},z,\qperp;y,\pperp) \dw(z,\qperp|M^2), 
\end{equation}
where we have 
\begin{equation} \label{F}
2 R^+ V_{\text{F}}(x,\kperp;y,\pperp;z,\qperp)^{-1}= R^- - \kminus_{\text{on}} - \frac{(\pperp - \kperp)^2 + \mu^2}{2 R^+(y-x)}- \frac{(\qperp - \pperp)^2 - \mu^2}{2 R^+(z - y)}- (R -q)^{-}_{\text{on}}.
\end{equation}

Having spelled out this last energy denominator which has two force carrying bosons simultaneously propagating, we refer the reader to Figure \ref{fv2} to observe that diagram F corresponds to Eq. \eqref{mess}. The remaining case in Table \ref{t:v2} is evaluated similarly to the case just worked. There are two residues picked up and hence an algebraic manipulation required to untangle diagram C and diagram E.  With all possible contributions to the second order wave function, we must now summarize the results thereby deriving the improved integral equation for the bound state. 

Each term we have evaluated has an overall factor of $\frac{1}{2R^+(\kminus_{b} - \kminus_{a})} = \dw(x,\kperp_{\text{rel}}|M^2)$ which sits outside of the
$p$ and $q$ integrals. There is also an overall factor of $\frac{1}{2R^+(\qminus_{b} - \qminus_{a})} = \dw(z,\qperp_{\text{rel}}|M^2)$ which multiplies
$\gamma(z,\qperp_{\text{rel}}|M^2)$---these can be combined to form $\psi^{(0)}(z,\qperp_{\text{rel}})$. Consider the terms leading to diagrams A-D of 
Figure \ref{fv2}. Writing these out explicitly (omitting an overall $\dw(x,\kperp_{\text{rel}}|M^2) \theta[x(1-x)]$ and factor of $\theta[y(1-y)] 
\theta[z(1-z)]$ in the integrals), we find\footnote{Notice the reversal in the order of the variables $p$ and $q$ in the interactions. This is only because we evaluated the $\pminus$ integral first which sat in the middle of Eq. \ref{so}. Such are the joys of notation.}
\begin{multline} \label{messy}
\int \frac{dy \; d\pperp}{2 (2\pi)^3 y(1-y)} \frac{dz \; d\qperp}{2 (2\pi)^3 z(1-z)} \Bigg( \theta(y-x) \tilde{V}(\kminus_{a},x,\kperp;y,\pperp) 
\dw(y,\pperp|M^2) \theta(y-z) \tilde{V}(\qminus_{a},z,\qperp;y,\pperp)\\ 
+ \theta(x-y) \tilde{V}(\kminus_{b},y,\pperp;x,\kperp) \dw(y,\pperp|M^2) \theta(z-y) \tilde{V}(\qminus_{b},z,\qperp;y,\pperp)\\ 
+ \theta(x-y) \tilde{V}(\kminus_{b},x,\kperp;y,\pperp) \dw(y,\pperp|M^2) \theta(y-z) \tilde{V}(\qminus_{a},z,\qperp;y,\pperp)\\  
+ \theta(y-x) \tilde{V}(\kminus_{a},x,\kperp;y,\pperp) \dw(y,\pperp|M^2) \theta(z-y) \tilde{V}(\qminus_{b},z,\qperp;y,\pperp) \Bigg) 
\psi^{(0)}(z,\qperp),
\end{multline}
which can be rearranged as
\begin{multline}
\int \frac{dy \; d\pperp}{2 (2\pi)^3 y(1-y)} \Big( \theta(x-y) \tilde{V}(\kminus_{b},x,\kperp;y,\pperp) + \theta(y-x) \tilde{V}(\kminus_{a},x,\kperp;y,\pperp) \Big)  \dw(y,\pperp|M^2) \\ 
\times \frac{dz \; d\qperp}{2 (2\pi)^3 z(1-z)} \Big( \theta(z-y) \tilde{V}(\qminus_{b},z,\qperp;y,\pperp) + \theta(y-z) \tilde{V}(\qminus_{a},z,\qperp;y,\pperp)  \Big) \psi^{(0)}(z,\qperp)\\
=  \int \frac{dy \; d\pperp}{2 (2\pi)^3 y(1-y)} V(y,\pperp;x,\kperp) \psi^{(1)}(y,\pperp) .  
\end{multline}
In the last line, we used the first step in the iteration scheme to relate $\psi^{(0)}$ to $\psi^{(1)}$ and the definition in Eq. \eqref{OBE}. 
At this point, the second order wave function depends on both the first order and zeroth order wave functions. To derive the bound state equation, 
we merely change all wave functions to second order since any difference is at least third order. Finally, we have derived the second-order bound 
state equation for the light-front wave function
\begin{equation}
\psi^{(2)}(x,\kperp) = \dw(x,\kperp|M^2) \int \frac{dy \; d\pperp}{2 (2\pi)^3 y(1-y)} \frac{dz \; d\qperp}{2 (2\pi)^3 z(1-z)} V(x,\kperp;y,\pperp;z,\qperp) 
\psi^{(2)}(z,\qperp),
\end{equation}
where the interaction $V$ to second order (and hence three sets of variables) is\footnote{This is of course the potential to second order in the ladder 
approximation. The full kernel contains additional second order pieces, such as a crossed ladder term, which contribute to the light front 
kernel beginning at second order.}
\begin{multline} \label{V2}
 V(x,\kperp;y,\pperp;z,\qperp) = 2 (2\pi)^3 y(1-y) \delta(z-y) \delta(\qperp - \pperp) V(x,\kperp;y,\pperp)\\
 + \theta(x-y) \tilde{V}(\kminus_{b},x,\kperp;y,\pperp) V_{\text{E}}(x,\kperp;y,\pperp;z,\qperp)\theta(y-z)\tilde{V}(\qminus_{a},z,\qperp;y,\pperp)\\
 + \theta(y-x)\tilde{V}(\kminus_{a},x,\kperp;y,\pperp) V_{\text{F}}(x,\kperp;y,\pperp;z,\qperp)
\theta(z-y)\tilde{V}(\qminus_{b},z,\qperp;y,\pperp),
\end{multline}
with $V(x,\kperp;y,\pperp)$ as the one-boson exchange potential \eqref{OBE} and $V_{F}$ as appears in Eq. \eqref{F}. The term $V_{E}$
which comes from diagram E is
\begin{equation} \label{E}
2 R^+ V_{\text{E}}(x,\kperp;y,\pperp;z,\qperp)^{-1}= R^- - \qminus_{\text{on}} - \frac{(\pperp - \qperp)^2 + \mu^2}{2 R^+(y-z)}- 
\frac{(\kperp - \pperp)^2 + \mu^2}{2 R^+(x - y)}- (R -k)^{-}_{\text{on}}.
\end{equation}

Now that we have found a scheme to derive the light front wave function from time ordered perturbation theory with a covariant starting
place, it is time to derive the scheme in a formal setting. This will enable us  more easily to consider matrix elements of current operators.

\section{Formal reduction} \label{formal}
In this section, we largely follow the reduction scheme set up in \cite{Sales:1999ec} to convert the covariant Bethe-Salpeter equation 
into a three-dimensional light front version using the results of section \ref{intuitive} to motivate the choice of auxiliary Green's function. 
We must first adjust our notation to be more in line with theirs, where needed. Above, we have removed overall
four dimensional, momentum-conserving delta functions, e.g. our propagator $G(k,R)$ is the momentum space version of $G(R)$ defined 
by $\langle R^\prime | G | R \rangle = (2 \pi)^4 \delta^{(4)}(R^\prime - R) G(R)$. Additionally we have removed the momentum conserving delta function 
between initial and final states of particle one in the disconnected Green's function, i .e.~ $ \langle k | G(R) | p \rangle = 
 (2\pi)^4 \delta^{(4)}(k - p) G(k,R)$. Using $\langle R^\prime | V | R \rangle = (2\pi)^4 \delta^{(4)}(R - R^\prime) V(R)$, we see the kernel $V(k,p)$
 defined above is $V(k,p) = \langle k | V(R) | p \rangle$. Notice the dependence on the total momentum $R$ drops out since we are specializing 
to the ladder approximation, \emph{cf} Eq. \eqref{ladder}. 
In terms of these fully four dimensional quantities, the Lippmann-Schwinger equation for the two-particle transition matrix $T$ appears as
\begin{equation} \label{LS}
T = V + V G T.
\end{equation}
A pole in the $T$-matrix (at some $R^2 = M^2$, say) corresponds to a two-particle bound state. Investigation of the pole's residue
gives the Bethe-Salpeter equation \eqref{Gamma} for the bound state vertex $\Gamma$
\begin{equation}
\Gamma = V G \Gamma.\notag
\end{equation}
The Bethe-Salpeter amplitude $\Psi$ is then $G \Gamma$. Above, our $\Psi_{\text{BS}}(k,R)$ is the momentum space version of $\Psi$ with
delta functions removed (as is similarly true of the vertex $\Gamma(k,R)$ which we implicitly used above in going from Eq. \eqref{Gamma} 
to Eq. \eqref{BS}). Following \cite{Sales:1999ec}, it is convenient to denote quantities able to be rendered in position or momentum space with 
bras and kets. We will employ this notation only for quantities that have been stripped of their overall momentum-conserving delta functions, 
for example $\Gamma(k,R) = \langle k | \Gamma_{R} \rangle $, where $R$ is used as a label for the bound state for which $R^2 = M^2$. 

\subsection{The scheme}
To reduce the Lippmann-Schwinger equation to a three-dimensional light front version, we must introduce an auxiliary Green's function\footnote{Of course,
the new Green's function must maintain the unitarity of the theory and thus is not completely arbitrary.}
$\Gt$ in place of $G$ (as in \cite{Woloshyn:wm}). To inject $\Gt$, we rewrite Eq. \eqref{LS} in the form
\begin{align}
T^{-1} & = V^{-1} - G \notag \\
       & = \underbrace{V^{-1} - (G - \Gt)}_{W^{-1}} - \Gt, 
\end{align} 
in which $V^{-1} - (G - \Gt)$ has the form of an inverse interaction (denoted $W^{-1}$) with respect to the auxiliary Green's function $\Gt$. 
Re-inverting $T$, we find that
\begin{equation}
T = W + W \Gt T,
\end{equation}
provided that
\begin{equation} \label{W}
W = V + V (G - \Gt) W. 
\end{equation}
This gives us an alternate way to arrive at the bound state vertex function 
\begin{equation} \label{regamma}
\Gamma = W \Gt \Gamma. 
\end{equation}
Until we choose $\Gt$ to reduce the Bethe-Salpeter
equation, we have only formally created a more complicated problem to solve.

To choose a light front reduction, $\Gt$ must inherently be related to projection onto the initial surface $x^+ = 0$. For simplicity, we denote
the integration $\int d\kminus \langle \kminus | \mathcal{O}(R) = \Big| \mathcal{O}(R)$. With this notation, we will always work in momentum space
for which the only sensible matrix elements of $\Big| \mathcal{O}(R)$ are of the form $\langle \kplus, \kperp| \;  \Big| \mathcal{O}(R) | 
\pminus, \pplus, \pperp \rangle$. The operator $\mathcal{O}(R) \Big|$ is defined similarly. 
For a useful reduction scheme (one that preserves unitarity), 
we must have $\Big| G(R) \Big| = \Big| \Gt(R) \Big|$. Bearing in mind an extra delta function in $G(R)$ relative to section \ref{intuitive}, we have already 
calculated $\Big| G(R) \Big|$ in Eq. \eqref{g} above, thus
\begin{equation}
\langle \kplus, \kperp | g(R) | \pplus, \pperp \rangle \equiv \langle \kplus, \kperp | \; \Big| \Gt(R) \Big| \; | \pplus, \pperp \rangle = 
\delta(\kplus - \pplus) \delta (\kperp - \pperp) \frac{2 \pi i \; \theta[x(1-x)]}{2 R^+ x(1-x)} \dw(x,\kperp_{\text{rel}}|R^2).
\end{equation}  
Furthermore, we wish to choose $\Gt$ in order that our reduction scheme reproduces time ordered perturbation theory. Yet in section 
\ref{intuitive} we were able to recover time ordered perturbation theory without explicit recourse to an auxiliary Green's function.
Therefore, we wish to keep $\Gt$ as close as possible to $G$. Referring back to the intuitive reduction scheme, (once we insert 
momentum conserving delta functions between the two particles) we realize that we must have tacitly used
\begin{align} \label{extra}
\Gt(R) \Big| & = G(R) \Big| \notag\\
\Big| \Gt(R) & = \Big| G(R).
\end{align}
This simplest way to achieve this requires 
\begin{equation} \label{gt}
\Gt(R) = G(R) \Big| g^{-1}(R) \Big| G(R),
\end{equation} 
for which the explicit mention of $\Gt$ disappears. We can then write the reduced $T$-matrix in the form
\begin{equation} \label{tred}
t(R) \equiv g^{-1}(R) \Big| \Gt(R) T(R) \Gt(R) \Big| g^{-1}(R) = g^{-1}(R) \Big| G(R) T(R) G(R) \Big| g^{-1}(R),
\end{equation}
where $t$ is a three-dimensional auxiliary transition matrix. As a function of the total momentum $R$, if $T$ has a bound state pole, so does $t$. 
This enables us to relate the four-dimensional wave function to a three-dimensional one. To this end, we write
\begin{equation} \label{lilt}
t(R) = w(R) + w(R) g(R) t(R),
\end{equation}
where the reduced auxiliary kernel is
\begin{equation}
w(R) = g^{-1}(R) \Big| G(R) W(R) G(R) \Big| g^{-1}(R).
\end{equation}
Taking the residue of Eq. \eqref{lilt} at $R^2 = M^2$, gives a homogeneous equation for the three-dimensional vertex function $\gamma$
\begin{equation} \label{gammaredu}
| \gamma_{R} \rangle = w(R) g(R) | \gamma_{R} \rangle,
\end{equation}
from which we can define the light front wave function $|\psi_{R}\rangle \equiv g(R) |\gamma_{R}\rangle$. 
By iterating the Lippmann-Schwinger equation for $T$ twice, it is possible to relate $T$ to $t$ and thereby construct $T$ given $t$, which is 
clearly not possible from the definition \eqref{tred}. Taking the residue of this relation between $T$ and $t$ yields the  
three-dimensional to four-dimensional conversion between bound state vertex functions, namely
\begin{equation} \label{convert}
| \Gamma_{R} \rangle = W(R) G(R) \Big| \;| \gamma_{R} \rangle.
\end{equation}
Lastly, we can manipulate the four-dimensional Bethe-Salpeter amplitude into the form
\begin{equation} \label{324}
| \Psi_{R} \rangle = \Bigg( 1 + \Big(G(R) - \Gt(R)\Big)W(R) \Bigg) G(R) \Big| \; |\gamma_{R} \rangle,
\end{equation}
which justifies the interpretation of $|\psi_{R}\rangle$ as the light front wave function since
\begin{equation}
\Big| \; |\Psi_{R}\rangle = \Bigg(  g(R) + \underbrace{\Big( \Big| G(R) - \Big| \Gt(R)\Big)}_{0}W(R) G(R) \Big|  \Bigg) |\gamma_{R} \rangle = 
|\psi_{R}\rangle.
\end{equation}
We comment that all light front reduction schemes when summed to all orders yield the 
$x^+ = 0$ projection of the Bethe-Salpeter equation. The result of this section is 
stronger since the kernel is calculated in light front time ordered perturbation theory,
e.g. $w(R)$ calculated to second order is identical to our result of section \ref{intuitive}
and carried out in this formal notation in \cite{Sales:1999ec}. Lastly the normalization of 
the covariant and reduced wave function is discussed in Appendix A. 

\subsection{Reversal of logic}
Having used the intuitive reduction scheme to motivate the form of the light front reduction 
found in \cite{Sales:1999ec}, we now reverse the logical circle and use the formal reduction to 
derive the iterative scheme of section \ref{intuitive}. 

The heart of the intuitive scheme lies in integrating out the minus momentum dependence of the
covariant wave function. So we merely rewrite this order-by-order in the formal reduction scheme.
Utilizing Eqs. \eqref{regamma} and \eqref{W} while omitting total four-momentum labels since they 
are all identical, we can write
\begin{align}
| \psi \rangle = \Big| \; | \Psi \rangle & = \Big| G W G \Big| g^{-1} |\psi\rangle \notag \\
				     & = \Big| G V \sum_{n = 0}^{\infty} \Big[ \Big( G - \Gt \Big) V \Big]^n G \Big| \;  |\gamma\rangle.
\end{align}
From truncating the series in $n$ at some $j < \infty$, we are led to a natural sequence of better approximations to the light front
wave function provided the interaction is weak. Working out the two lowest orders to parallel the discussion above, we have
\begin{equation}
|\psi^{(1)}\rangle = \Big| G V G \Big| \; |\gamma\rangle,
\end{equation}
which looks like one iteration of the Bethe-Salpeter equation and
\begin{align}
|\psi^{(2)}\rangle & = \Bigg ( \Big| G V G \Big| + \Big| G V \Big(G - \Gt \Big) V G \Big|  \Bigg) |\gamma\rangle \notag \\
                   & = \Big| G V G V G \Big| \; |\gamma\rangle,
\end{align}
(after having used Eqs. \eqref{regamma} and \eqref{convert} to get to the second line) looks like two iterations of the Bethe-Salpeter equation.
One might wonder how the instantaneous approximation appears in any of this. Using again Eqs. \eqref{regamma} and \eqref{convert}, we can show
\begin{equation} \label{key}
\Gt \Big| \Gamma \rangle = G \Big| \; |\gamma\rangle
\end{equation}  
and notice
\begin{align}
\Big| G V G \Big| \; |\gamma\rangle & = \Big| G V \Gt | \Gamma \rangle \notag \\
\Big| (G V)^2 G \Big| \; |\gamma \rangle & = \Big| (G V)^2 \Gt | \Gamma \rangle \notag \\
	\begin{CD}
 	@VVV
	\end{CD} \notag \\
\Big| (G V)^j G \Big| \; |\gamma \rangle & = \Big| (G V)^j \Gt | \Gamma \rangle. \notag
\end{align}
Thus at any order in the scheme of section \ref{intuitive} it is as if we have iterated $j$ 
times to arrive at $(GV)^j G |\Gamma \rangle$ and replaced the lone $G$ to the immediate left of $\Gamma$ with $\Gt$. 
The instantaneous approximation then requires $\Gt |\Gamma \rangle$ to have no minus momentum poles besides those of the propagator 
which is just the statement made by equation \eqref{key}.  

With a formal reduction (that agrees with our intuitive scheme) in hand, we are now ready 
to tackle computing matrix elements of currents between bound states and related applications. In turn, these applications
will help us to grasp better the nature of the reduction scheme.

\section{Current in the reduced formalism}\label{current}
In this section, we extend the formalism presented so far to include current matrix elements between bound states. We do so in a gauge invariant 
fashion following \cite{Gross:bu}. Our notation, however, is more in line with the elegant method of gauging equations presented in 
\cite{Kvinikhidze:1998xn}. This latter method is far more general and extends to bound systems of more than two particles. 

\begin{figure}
\begin{center}
\epsfig{file=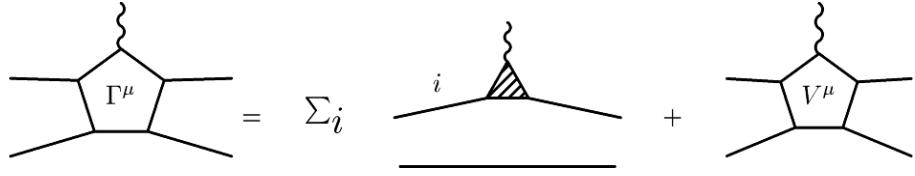}
\end{center}
\caption{Graphical depiction of the irreducible five-point function $\Gamma^\mu$ as 
sum of impulse terms and a gauged interaction. By construction, 
$\Gamma^\mu$ is gauge invariant.}
\label{fgamma}
\end{figure}

Consider first the full four-point function $G^{(4)}$ defined by
\begin{equation}\label{g4}
G^{(4)} = G + G T G. 
\end{equation}
For later use, it is important to note that the residue of $G^{(4)}$ at the bound state pole $R^2 = M^2$ is $- i |\Psi_{R}\rangle\langle \Psi_{R}|$. Using the Lippmann-Schwinger equation for $T$, we can show the four-point function
satisfies 
\begin{equation}\label{LSg4}
G^{(4)} = G + G V G^{(4)}. 
\end{equation}

To discuss electromagnetic current matrix elements, we will need the three-point funtion
function $d_{i}^\mu$ where the label $i$ denotes particle number (the utility will be clear momentarily). Now we define an irreducible
three-point function $\Gamma_{i}^\mu$ in the obvious way
\begin{equation}
d_{i}^\mu = d_{i} \Gamma_{i}^\mu d_{i}.
\end{equation}
Now we need to relate the one-particle electromagnetic vertex function to the $T$ matrix. Let $d^\mu \equiv \overset{\leftrightarrow}\partial{}^\mu$ denote the electromagnetic coupling to our scalar particles. Since the electromagnetic three-point function $\Gamma_{i}^\mu$ is irreducible, we have
\begin{equation}
\Gamma_{i}^\mu  = G^{-1} G^{(4)} d^\mu 
\end{equation}
 and by using the definition of $G^{(4)}$ Eq. \ref{g4}, we have the desired relation
\begin{equation}
\Gamma_{i}^{\mu} = d^\mu + T G d^\mu.
\end{equation}
Notice the right hand side lacks the particle label $i$. In the first term, the bare coupling acts on the $i$th paticle while in the second term the bare coupling does not act on the $i$th particle..
For this reason we have dropped the label which will always be clear from context.

In considering two propagating particles' interaction with a photon, the above definitions lead us to the impulse approximation
to the current
\begin{equation}
\Gamma_{0}^\mu = \Gamma_{1}^\mu d_{2}^{-1} + d_{1}^{-1} \Gamma_{2}^\mu.
\end{equation}
Additionally the photon could couple to the particles in the midst of their interacting. Define a gauged interaction $V^\mu$ topologically
by attaching a photon to the kernel in all possible places. This leads us to the irreducible electromagnetic vertex $\Gamma^\mu$ defined as (see Figure \ref{fgamma})
\begin{equation} \label{emvertex}
\Gamma^\mu = \Gamma_{0}^\mu + V^\mu,
\end{equation}
which is gauge invariant by construction.

\begin{figure}
	\begin{center}
	\epsfig{file=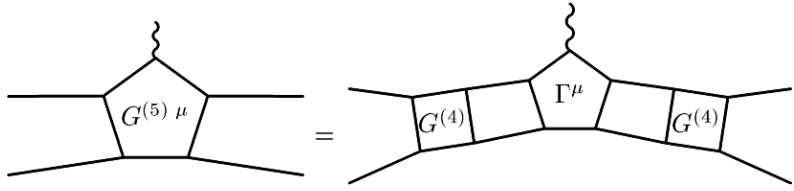}
		\end{center}
	\caption{Graphical depiction of the five-point function $G^{(5) \; \mu}$. The 
	irreducible five-point function is the gauge invariant $\Gamma^\mu$.}
	\label{f5pt}
\end{figure}

Lastly to calculate matrix elements of the current between bound states it is useful to define a reducible five-point function
(see Figure \ref{f5pt})
\begin{equation} \label{5alive}
G^{(5) \; \mu} = G^{(4)} \Gamma^\mu G^{(4)}.
\end{equation}

Having laid out the necessary facts about electromagnetic vertex functions and gauge invariant currents, we can now specialize to their matrix elements between 
bound states.

\subsection{Form factors at leading order}
Consider Figure \ref{f5pt}. Knowing that the four-point function $G^{(4)}$ (by way of the analytic structure of $T$) has a bound state pole, we can easily deduce the current between bound states. It is clear from the figure, that electromagnetic scattering between free two-particle states is described by the current $J_{oo}^\mu = G^{-1} G^{(5)\;\mu} G^{-1}$. From this it is easy to generalize to scattering between bound states (labeled by $P,i=$ initial and $P^\prime,f=$ final) by taking the appropriate residues of the five-point function, namely
\begin{equation}
J_{fi}^\mu \propto \Res\Big(G^{(5)\;\mu};P^2 = M^2_{i}, P^\prime{}^2= M^2_{f}\Big). \notag
\end{equation}
Equality is achieved by removing $|\Psi_{P^\prime}\rangle$ and $\langle \Psi_{P} |$. By using Eq. \eqref{5alive}, we can cast the current into the form
\begin{equation} \label{currentme}
J_{fi}^\mu = \langle \Psi_{P^\prime} | \Gamma^\mu | \Psi_{P} \rangle.
\end{equation}
The vertex function satisfies the Ward-Takahashi identity, so that one can easily use the Bethe-Salpeter equation for $\Psi$ to demonstrate 
current conservation: $(P^\prime - P) \cdot J_{fi} = 0$. 

We now specialize to the case where the initial and final states are the same particle. 
Lorentz covariance and current conservation dictate\footnote{Additionally
in the light front formalism, there can be an additional spurious form factor $B(t)$. This is due to the presence of an extra four-vector in the problem
$\omega^\mu$, which we introduced by needing to specify the orientation of the light front plane. For a composite scalar, $B(t)$ can be removed simply
by looking at the $+$ component of the current $J^\mu$ \cite{Carbonell:1998rj}. This is a manifestation of the linearity of the angular condition for 
scalar particles.} 
\begin{equation} \label{formula}
J^\mu = - i (P^\prime{}^\mu + P^\mu) F(t),
\end{equation}
in which appears the electromagnetic form factor $F(t)$, with $t = (P^\prime - P)^2$. 
In this section, we calculate the form factor to leading order in perturbation theory. Typically one chooses the $+$ component of the current as well as a frame in which $\Delta^+ = 0$ (where $\Delta$ is defined by $\Delta^\mu =  P^\prime{}^\mu - P^\mu$). As we plan to focus on nonvalence contributions as well as generalized parton distributions, we do not choose $\Delta^+ = 0$ which puts us in a situation similar to Sawicki \cite{Sawicki:1991sr}, who considered such expressions earlier. We do, however, choose to work in a frame which simplifies the transverse momentum dependence, namely one in which $\mathbf{P}^\perp = 0$. 

\begin{figure}
 \begin{center}
\epsfig{file=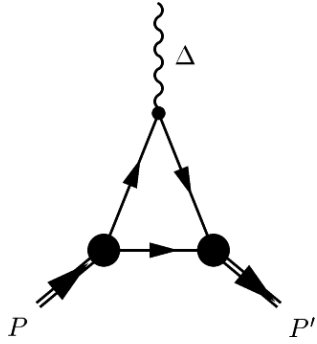}
  \caption{The leading order diagram for the electromagnetic form factor}
\label{ftri}
\end{center}
\end{figure}

At leading order, we can remove any explicit dependence upon the interaction from the current, thus $V^\mu$ can be neglected. Similarly $\Gamma_{i}^\mu = d^\mu + T G d^\mu \approx d^\mu$ since the Born series for $T$
starts off at $V$. Since our bound particles are equally massive scalars there is no essential difference in attaching the photon to particle one versus particle two. Hence at leading order we have
\begin{equation}\label{LO}
J_{\text{LO}}^\mu = 2 \langle \Psi_{P^\prime} | d^\mu d_{2}^{-1} | \Psi_{P} \rangle. 
\end{equation}
Instead of calculating the covariant form presented above, we use the results of section \ref{formal} to express the leading order form factor in terms of the light front wave function, via equation \eqref{324}. But to leading order, 
$|\Psi\rangle \approx G \Big| \; |\gamma \rangle$. Using the plus component of the current results in
\begin{equation} \label{formfactor}
F(t) = \frac{i}{P^+(1 - \zeta/2)} \langle \gamma_{P^\prime} | \; \Big| G(P^\prime) d^+ d_{2}^{-1} G(P) \Big| \; | \gamma_{P} \rangle,
\end{equation}  
where we have defined $\zeta = - \Delta^+ / P^+$. Inserting the effective resolution of unity, we can convert the above expression into
\begin{equation}
F(t) = \frac{i}{P^+(1-\zeta/2)} \int \frac{d^4 p}{(2\pi)^4} \; \frac{d^4k}{(2\pi)^4} \langle \gamma_{P^\prime} | \pplus,\pperp \rangle \; \langle
p | G(P^\prime) d^+ d_{2}^{-1} G(P) | k \rangle \; \langle \kplus,\kperp | \gamma_{P} \rangle.
\end{equation}
This expression simplifies since the disconnected Green's function still possesses an overall delta function
$\langle p | G(P) | k \rangle = (2\pi)^4 \delta^{(4)}(p -k ) G(k,P)$. After further insertions of unity, we have the factor
\begin{equation}
\langle p | d^+ | k \rangle = -i (2 \kplus + \Delta^+) (2\pi)^4 \delta^{(4)}(p - k - \Delta).  
\end{equation}
Now define $x = \kplus/P^+$. With this parameterization and performing the last integral rendered trivial by the above delta function, the expression for 
the form factor reads (see Figure \ref{ftri})
\begin{equation} \label{formLO}
F(t)  = \frac{1}{1-\zeta/2} \int \frac{d^4 k}{(2 \pi)^4}  \gamma^*(y,\kperp + (1-y) \mathbf{\Delta}^\perp |M^2)  d(k + \Delta) 
\big( 2 x - \zeta \big) d(k) d(P-k) \gamma(x,\kperp |M^2),
\end{equation}
where for the final meson's reduced vertex, we have employed $y = \frac{\kplus + \Delta^+}{P^\prime{}^+} = \frac{x- \zeta}{1-\zeta}$. Since the vertex functions are reduced, there is only $\kminus$ dependence in the propagators which enables integration. The poles are situated at
\begin{equation}
\begin{cases}
	\kminus_{a} = \kminus_{\text{on}} - \frac{\ie}{x}\\
	\kminus_{b} = P^- + (k-P)^-_{\text{on}} - \frac{\ie}{x-1}\\
	\kminus_{c} = - \Delta^- + (k + \Delta)^-_{\text{on}} - 
\frac{\ie}{x-\zeta} 
\end{cases}
\end{equation}
We cannot assume, however that contributions are only for $0<x<1$ and $0<y<1$, since the form factor appears as a matrix element of reduced vertices rather than light front wave functions. We went to some labor to show that the wave function has support only between zero and one. The reduced vertex need not vanish for $x$ outside of zero to one, rather utilizing Eq. \eqref{gammaredu} 
\begin{align} \label{gammared}
\gamma(y, \pperp |M^2) & = \langle y R^+, \pperp | \gamma_{R} \rangle \notag \\
		       & = \int \frac{dx d\kperp}{2(2\pi)^3 x(1-x)} w(y,\pperp; x,\kperp|M^2) \dw(x, \kperp|M^2) \gamma(x, \kperp |M^2) 
\end{align}
demonstrates that $\gamma$ for $0<x<1$ uniquely determines $\gamma$ everywhere. 

Evaluation of the integral over $\kminus$ in Eq. \eqref{formLO} is straightforward (and carried out in \cite{Tiburzi:2001ta}). We pick up only contributions from $0<x<1$, $y$ on the other hand is between $\frac{-\zeta}{1-\zeta}$ and one. There are two distinct regions of the integration, $0 < x <\zeta$ and $\zeta< x < 1 $ (or equivalently, $\frac{-\zeta}{1-\zeta} < y < 0$ and $0< y < 1 $ respectively) which correspond to picking up different residues. For $x>\zeta$, the time ordered diagram which we pick up corresponds to Figure \ref{ftri} (which has the luck of being covariant on one occasion and now time ordered). And in this case, since both $x$ and $y$ are between zero and one, both the initial and final state vertices can be described by a light front wave function. Evaluation yields
\begin{equation}\label{FFLO}
F(t) = \frac{1}{1 - \zeta/2} \int \frac{dx d\kperp}{2 (2\pi)^3} \frac{2 x - \zeta}{x y (1-x)} \psi^*\big(y,\kperp + (1 - y) \mathbf{\Delta}^\perp \big) \psi\big(x,\kperp\big).
\end{equation}
Contribution from this region survives the limit $\zeta \to 0$. Contribution from the other region of integration vanishes in this limit (modulo possible zero mode singularities). In this limit, which corresponds to the usual frame for the evaluation of light front form factors, Eq. \eqref{FFLO} reproduces the Drell-Yan formula for the electromagnetic form factor \cite{Drell:1970km}. 

\begin{figure}
	 \begin{center}
	\epsfig{file=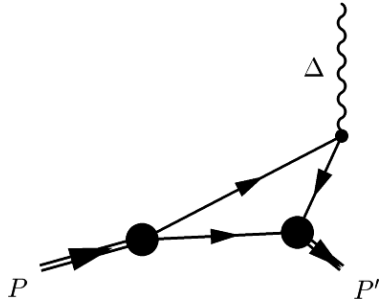}
 	\caption{The $Z$ graph which confronts us to leading order in evaluating the electromagnetic form factor}
 	\label{fZ}
 	\end{center}
\end{figure}

In the other region $x<\zeta$, however, a $Z$ graph confronts us in extracting the form factor, see Figure \ref{fZ}. As in our previous work, we arrive at an expression
\begin{equation} \label{evil}
F(t) \sim \int \gamma^*(y,\ldots) g \psi(x,\ldots) \notag
\end{equation}
where to define the vertex (since $y<0$) we must appeal to Eq. \eqref{gammared}. If interpreted graphically, we have fixed up the non-wave function vertex by appealing to crossing symmetry. This crossed interaction, however, involves a 
time ordering with a parton traveling back through time, and ignores the spectator of momentum $k$. As precarious as it sounds, this corresponds to hadronization off the quark line and is precisely what the formal reduction dictates---at least from iterating the reduced Bethe-Salpeter equation. But there is obviously a reason we have not spelled out the exact functional form of this contribution. Here we are ascertaining the form factor at leading order. Using Eq. \eqref{gammared} to cross the interaction, introduces a power of the weak coupling and thus this nonvalence contribution is at a higher order than which we are currently working.\footnote{By using Eq. \eqref{gammared} for the vertex, one might wonder how the interaction is small since $g w \sim 1$ for the bound state. The key is that we are dealing with a non-wave function vertex. Having crossed the interaction, we are describing a hadronization potential which 
cannot be reduced into the final state wave function.} To be correct, the region $0< x < \zeta$ then contributes nothing at leading order, but leaves a puzzle for the first order corrections.

\subsection{Form factors at next to leading order}
Above we have derived the leading order expression for the electromagnetic form factor. We were confronted with a nonvalence contribution, which we temporarily dealt with by iterating the reduced Bethe-Salpeter equation \eqref{gammared}. The process of iteration, however, launched us into a higher order of perturbation theory and thus as far as leading order expressions are concerned, the contribution from $0< x< \zeta$ vanishes. Here we work at next to leading order in perturbation theory to unravel the valence and nonvalence contributions to the form factor.

At leading order in the weak coupling, we must retain the first term in the Born series as well as the leading term in converting from the Bethe-Salpeter wave function to the reduced vertex. Again there is no contribution from the gauged interaction as it enters at the next order. Referring to equation \eqref{currentme}, we have
\begin{align}
J^\mu & \approx 2 \langle \gamma_{P^\prime} | \; \Big| G(P^\prime) \Big( 1 + V(P^\prime) ( G(P^\prime) - \Gt(P^\prime)) \Big)\Big( d^\mu d_{2}^{-1} + V(-\Delta) G(-\Delta) d^\mu d_{2}^{-1} \Big) \Big( 1 + (G(P) - \Gt(P)) V(P)  \Big) G(P) \Big| \; | \gamma_{P} \rangle   \notag \\
	& = J^\mu_{\text{LO}} + \delta J^\mu_{i} + \delta J^\mu_{f} + \delta J^\mu_{\gamma} + \mathcal{O}[V^2],
\end{align}
in which appears the leading order result \eqref{LO} as well as the first order correction terms given by
\begin{equation} \label{NLO}
	\begin{cases}
		\delta J^\mu_{i} = 2 \langle \gamma_{P^\prime} | \; \Big| G(P^\prime) d^\mu d_{2}^{-1} \Big(G(P) - \Gt(P) \Big) V(P) G(P) \Big| \; 
|\gamma_{P} \rangle\\
		\delta J^\mu_{f} = 2 \langle \gamma_{P^\prime} | \; \Big| G(P^\prime) V(P^\prime) \Big(G(P^\prime) - \Gt(P^\prime) \Big) d^\mu d_{2}^{-1} 
G(P) \Big| \; | \gamma_{P} \rangle\\
		\delta J^\mu_{\gamma} = 2 \langle \gamma_{P^\prime} | \; \Big| G(P^\prime) \Big(V(-\Delta) G(-\Delta) d^\mu\Big) d_{2}^{-1} G(P) \Big| \; 
| \gamma_{P} \rangle.
	\end{cases}
\end{equation}
The notation is reminiscent of the intuitive reduction scheme of section \ref{intuitive}, e.g. $\delta J^\mu_{f}$ is the correction term due to iterating the Bethe-Salpeter equation for the final state once and then making an instantaneous vertex approximation (modulo two-particle reducible contributions),  while $\delta J^\mu_{\gamma}$ is the correction from retaining one order of the four-dimensional Born series. Indeed we could have derived the 
above form using the intuitive scheme. Our task is now clear: perform the minus 
momentum integrals to ascertain what contributions the new poles introduced by iteration will make.

\subsubsection{Born term}

Naturally, we opt to consider the simplest first. Let us write out the momentum representation of $\delta J^+_{\gamma}$ appearing in Eq. \eqref{NLO},
\begin{multline}
\delta J^+_{\gamma} = 2 i \int \frac{d^4 p}{(2\pi)^4} \frac{d^4k}{(2\pi)^4}  \gamma^*(v,\pperp + (1 - v) \mathbf{\Delta}^\perp |M^2)  
d(p+ \Delta) d(P-p) \\
\times V(p,k) d(k) d(k+\Delta) (-i) \Big( 2\kplus + \Delta^+\Big) d(p) \gamma(w,\pperp | M^2) ,
\end{multline}
where $w = \pplus/P^+$ and  $ v = \frac{w - \zeta}{1 - \zeta}$. Part of the $\kminus$ integral should be attacked first since it vanishes. We 
notice the $\kminus$ poles $\kminus_{a}, \kminus_{c}$, and $\kminus_{v}(\qminus)$ all have negative imaginary parts for $\zeta < w < x$ and hence 
the contribution to the form factor vanishes in this region. Now we can tackle the $\pminus$ integral with ease. Since the interaction 
is a correction to the bare photon vertex, this boson exchange can never be reduced into either vertex via the bound state equation. 
Thus we must have an explicit power of the weak coupling in any contributing diagrams. Were there non-wave function vertices to consider from the 
initial or final state, crossing symmetry would dictate they are at a higher order and can be neglected here. Hence we have both $0<w<1$ 
(from the initial vertex) and $0<v<1$ (from the final vertex). Combining the two inequalities, we have the constraint $\zeta <w< 1$. 
Furthermore, our above consideration of the $\kminus$ integral tells us that only $w > x$ is relevant. These last two inequalities fix the 
signs of the $\pminus$-poles' imaginary parts and the integral is $+2 \pi i \Res(\pminus_{b})$. Taking this residue produces the energy denominators
$\pminus_{b} - \pminus_{a}$, $\pminus_{b} - \pminus_{c}$ which turn the initial and final vertices into wave functions respectively and
$\pminus_{b} - \pminus_{v_{k}} \equiv \kminus_{v_{b}} - \kminus$. 

\begin{figure}
 \begin{center}
 \epsfig{file=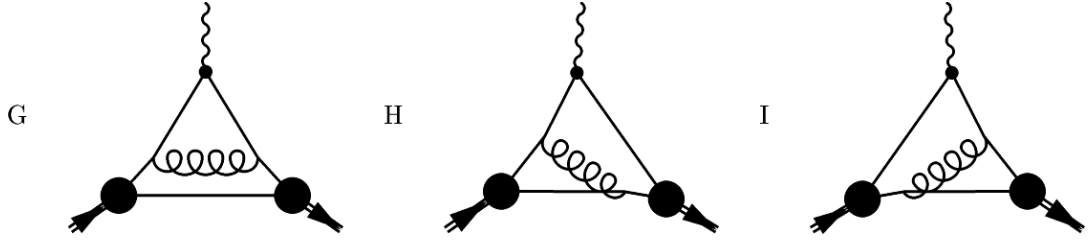}
\caption{Diagrams which contribute to the form factor at next to leading order (for $x >\zeta$).}
\label{ftri2}
\end{center}
\end{figure}

\begin{figure}
	 \begin{center}
\epsfig{file=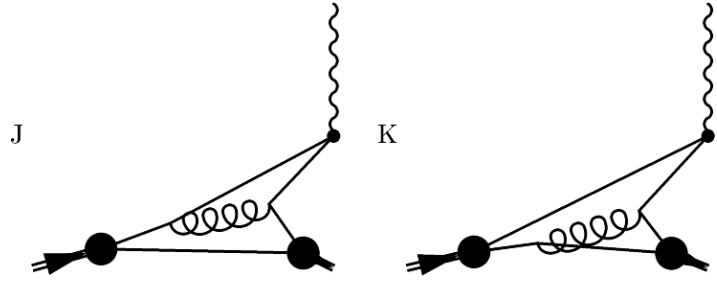}
 	\caption{The remaining diagrams (characterized by $x < \zeta$) for the electromagnetic 
		form factor at next to leading order.}
 	\label{fZZZ}
 	\end{center}
\end{figure}

Now in performing the $\kminus$ integration, we note that since $\zeta<w<1$ and $w>x$, we must have $x<1$. Note further, for $x<0$ all poles $\{ \kminus_{a}, \kminus_{c}, \kminus_{v_{b}} \}$ have positive imaginary parts. Hence we pick up only two contributions: $-2 \pi i \Res(\kminus_{a})$, for $0<x<\zeta$; and
$+2 \pi i \Res(\kminus_{v_{b}})$, for $\zeta<x<1$. Writing these out, we have for $\zeta<x<1$
\begin{multline} \label{bt1}
\frac{-2 i P^+}{[2(2\pi)^3]^2} \int \frac{dx d\kperp \; \theta(x - \zeta)}{x (x-\zeta)} \frac{dw d\pperp \; \theta(w - x)}{v w (1-w)} 
\; \frac{x - w}{g^2} \psi^*(v, \pperp + (1-v)\mathbf{\Delta}^\perp)\\
\tilde{V}(\kminus_{c},x,\kperp; w,\pperp) \Big( 2 x - \zeta  \Big) \tilde{V}(\kminus_{a}, x, \kperp ; w, \pperp) \psi(w,\pperp)
\end{multline}
where the energy denominators are given by
\begin{align}
\frac{g^2}{2 P^+(w - x)} \theta(w - x) \tilde{V}(\kminus_a,x,\kperp;w,\pperp)^{-1} 
				   & = P^- - \kminus_{\text{on}} - \frac{(\pperp - \kperp)^2 + \mu^2}{2P^+(w-x)} - (P - p)^-_{\text{on}}  ,\\
\frac{g^2}{2 P^+(w - x)} \theta(w - x) \tilde{V}(\kminus_{c},x,\kperp;w,\pperp)^{-1} 
				   & = P^- + \Delta^- - (k+\Delta)^-_{\text{on}} - \frac{(\pperp - \kperp)^2 + \mu^2}{2P^+(w-x)} - (P - p)^-_{\text{on}},
\end{align}
and for $0<x<\zeta$
\begin{multline} \label{bt2}
\frac{+ 2 i P^+}{[2 (2\pi)^3]^2} \int \frac{ dx d\kperp \; \theta(\zeta - x)}{ \sigma(1-\sigma) \zeta} \frac{dw d\pperp \; \theta(w-x)}{v w (1 - w)}  \psi^*(v,\pperp + (1-v)\mathbf{\Delta}^\perp)
\dw (\sigma, \kperp + \sigma \mathbf{\Delta}^\perp |\Delta^2) \\
\times \Big( 2 x - \zeta \Big) \tilde{V}(\kminus_{a},x,\kperp;w,\pperp) \psi(w,\pperp),
\end{multline}
where we have defined $\sigma = x/\zeta$. 
Each of these contributions can be interpreted diagrammatically. For $\zeta < x < 1$, we have the time ordered graph G of Figure \ref{ftri2}, and
for $0<x<\zeta$, we have the Z graph labeled J in Figure \ref{fZZZ}.  

\subsubsection{Initial state iteration}

At the next level of difficulty, let us write out the momentum 
representation of $\delta J^+_{i}$ appearing in Eq. \eqref{NLO},
\begin{multline}
\delta J^+_{i} = 2 i \int \frac{d^4 k}{(2\pi)^4} \frac{d^4 p}{(2\pi)^4}
\gamma^*(y,\kperp + (1-y) \mathbf{\Delta}^\perp |M^2)
d(P-k) d(k+\Delta) d(k) \\ 
\times (-i) \Big(2 \kplus + \Delta^+ \Big) V(k,p) d(p) d(R-p) \gamma(w,\pperp |M^2),  
\end{multline}
where again $w = \pplus/P^+$. For ease, we have left off the subtraction of the 
two-particle reducible contribution (arising from $\Gt$ defined in equation \ref{gt}). We shall need to make use of its form below and thus explicitly the 
missing term is
\begin{equation} 
- 2 \langle \gamma_{P^\prime} |\; \Big| G(P^\prime) d^\mu d_{2}^{-1} \Gt(P) V(P) G(P) \Big| \; | \gamma_{P} \rangle  = - 2 \langle \gamma_{P^\prime} 
| \; \Big| G(P\prime) d^\mu d_{2}^{-1} G(P) \Big| \; | \gamma_{P} \rangle. \notag
\end{equation} 
The form of this term is identical to Eq. \eqref{LO}, except with the crucial sign difference. Thus in evaluating $\delta J^+_{i}$, we shall merely throw away pieces that can be reduced back into the initial state wave function.  To perform the $\pminus$ integral, we note that $w$ is restricted to $0<w<1$ at this order. If there were contributions outside this range, we must appeal to crossing symmetry for the initial vertex, which necessarily introduces higher order terms. 

In carrying out the $\pminus$ integral, there are three poles: $\pminus_{a},
\pminus_{b}$ and $\pminus_{v}(\kminus)$. Given $0<w<1$, the first has a negative
imaginary part, the second positive and the third depends upon the sign of $x - w$. Taking the relevant residues produces an answer we are familiar with
\begin{equation} \label{midstep}
\int d\pminus \frac{\gamma(w,\pperp|M^2) V(p,k)}{(\pminus - \pminus_{a}) ( \pminus - \pminus_{b})} \propto \psi(w,\pperp) \tilde{V}(\kminus,x,\kperp;w,\pperp), 
\end{equation}
referring back to Eq. \eqref{vtilde} for the definition of $\tilde{V}$ (with the necessary renaming of variables). The resulting poles of the $\kminus$ integrand are: $\kminus_{a}, \kminus_{b}, \kminus_{c}, \kminus_{v_{a}}$ and $\kminus_{v_{b}}$, and there are two cases to consider buried in $\tilde{V}$ depending on the sign of $x - w$. The $\kminus$ integral is nearly the same as in section \ref{intuitive}, however, there is an additional pole $\kminus_{c}$.

Consider first, the term which results when $x > w$. In the case $x > \zeta$ we do not need to tackle a Z graph contribution and so we proceed along this course first. With these restrictions on the plus momentum fractions, the signs of the poles' imaginary parts are fixed and we may perform the integration by closing the contour in the upper-half plane, which produces $+ 2\pi i \Res(\kminus_{b})$. Recognizing the multiplicative factor
\begin{equation}
\kminus_{b} - \kminus_{c} = P^- + \Delta^- - (P - k)^-_{\text{on}}- (k + \Delta)^-_{\text{on}}
\end{equation}
allows us to convert the final state vertex into a wave function. The interaction dependent term is $\kminus_{b} - \kminus_{v_{a}}$ which is a time ordered one-boson exchange. This contribution then appears as
\begin{multline} \label{redme}
- 2 i P^+ \int \frac{dx d\kperp \; \theta(x - \zeta)}{2 (2\pi)^3 y x  (1-x)} \psi^*(y, \kperp + (1 -y) \mathbf{\Delta}^\perp)  \Big( 2 x - \zeta  \Big)   \dw(x,\kperp | M^2)\\
\times \int \frac{dw d\pperp \; \theta(x-w)}{2 (2 \pi)^3 w (1-w)}  V(x, \kperp; w, \pperp) \psi(w, \pperp) 
\end{multline}
and will be removed by an identical term stemming from $\Gt$ when combined with the other one-boson exchange uncovered below. 

For $x < w$, we take up the same restriction concerning the final-state vertex $x > \zeta$. The poles of the $\kminus$ integrand are: $\kminus_{a}, \kminus_{b},\kminus_{c}$ and $\kminus_{v_{b}}$. Once again, the signs of their imaginary parts are completely specified. Evaluation of the integral by closing the contour in the lower-half plane produces $-2 \pi i \Big( \Res(\kminus_{a}) + \Res(\kminus_{c}) \Big)$, for which any individual residue does not correspond to a time ordered graph. Their sum, however, can be rearranged
\begin{multline} \label{algebra}
\frac{1}{(\kminus_{a} - \kminus_{b})(\kminus_{a} - \kminus_{c})(\kminus_{a} - \kminus_{v_{b}})} + \frac{1}{(\kminus_{c} - \kminus_{a})(\kminus_{c} - \kminus_{b})(\kminus_{c} - \kminus_{v_{b}})} \\ = \frac{1}{(\kminus_{b} - \kminus_{a})(\kminus_{a} - \kminus_{v_{b}})(\kminus_{c} - \kminus_{b})}  - \frac{1}{(\kminus_{a} - \kminus_{v_{b}})(\kminus_{c} - \kminus_{b})(\kminus_{c} - \kminus_{v_{b}})}
\end{multline} 
into two time ordered contributions. The first term above corresponds to the other one-boson exchange then free propagation to interact with the photon and finally absorption into the final state; in essence, it is equation \eqref{redme} only with $\theta(w - x)$. Summing these two contributions allows us to utilize the bound state equation and the resulting expression is identical to the leading order form factor (for $x>\zeta$) Eq. \eqref{LO}. Hence these contributions are removed by $\Gt$ appearing in Eq. \eqref{NLO}. 

The second term above \eqref{algebra}, however, is not removed by $\Gt$. Assembling all the pieces evaluated by residues leads to
\begin{multline} \label{is1}
\frac{- 2 i P^+}{[2(2\pi)^3]^2} \int \frac{dx d\kperp \; \theta(x - \zeta)}{y x (1-x)} \; \frac{dw d\pperp \; \theta(w - x)}{w (1-w)} \; \frac{x - w}{g^2} \psi^*(y,\pperp + (1 - y) \mathbf{\Delta}^\perp) \tilde{V}(\kminus_{c},x,\kperp;w,\pperp)\\
\times  \Big( 2 x - \zeta \Big) \tilde{V}(\kminus_{a}, x, \kperp; w,\pperp) \psi(w,\pperp),
\end{multline}
which is graphically depicted by diagram H in Figure \ref{ftri2}.

Lastly in dealing with $\delta J^\mu_{i}$ we must consider the case $x<\zeta$ and come to terms with the Z graph contribution encountered at leading order. 
Before toiling away at the integrals, we note: the final state vertex will require crossing and thus the interaction present from iterating
the Bethe-Salpeter equation for the initial state must be reduced into the initial state vertex, else we are at higher order, which is beyond
our concern. Let us spell this out to be certain. 

To utilize the bound state equation to remove the interaction at the initial state, we need $0<w<1$ and $x>0$. Performing the $\pminus$ integral yields 
equation \eqref{midstep} above. Not surprisingly, there are two cases to consider depending on the sign of $w-x$. When $w>x$, only $\kminus_{a}$ has 
a negative imaginary part. Evaluating the residue gives us half of the one-boson exchange $(\kminus_{a} - \kminus_{v_{b}})^{-1}$ acting on the initial 
state wave function. On the other hand, when $x<w$ both $\kminus_{b}$ and $\kminus_{c}$ have positive imaginary parts. Taking the residue at $\kminus_{c}$
puts the boson exchange along the $q \bar{q}$ pair's path to be annihilated into the photon and hence cannot be absorbed into the initial state. The
remaining parts of the diagram require partons moving backward through time and the crossed vertex. At next to leading order, this contribution is
thus zero. Evaluating the residue at $\kminus_{b}$ produces the factor $(\kminus_{b} - \kminus_{v_{a}})^{-1}$  which gives the other half of the 
one-boson exchange potential acting on the initial state wave function. Combining the two exchanges allows the interaction to be absorbed into the initial
state wave function. Thus we arrive at Figure \ref{fZ}. But this is precisely the leading order contribution to the form factor for $x < \zeta$ and is 
subsequently removed by $\Gt$. This is the key observation of this work. In light cone time ordered perturbation theory, appealing to crossing symmetry to deal 
with non-wave function vertices is an ineffectual fix. When one works correctly to a given order, the crossed interactions are removed.\footnote{ 
There is, however, the special case of an instantaneous interaction which we consider in Appendix B.}

\subsubsection{Final state iteration}
One final contribution to the form factor at next to leading order remains, $\delta J^\mu_{f}$ which arises from iterating the Bethe-Salpeter equation for 
the final state. Referring back to Eq. \ref{NLO}, we write this contribution in momentum space as
\begin{multline} \label{jeff}
\delta J^+_{f} = 2 i \int \frac{d^4 k}{(2 \pi)^4} \frac{d^4q}{(2\pi)^4} 
\gamma^*(z,\qperp - z \mathbf{\Delta}^\perp) d(q) d(P^\prime - q) V(q, k+ \Delta)\\
\times d(k+ \Delta) d(P - k) (-i) \Big(2 \kplus + \Delta^+ \Big) d(k) \psi(x,\kperp),
\end{multline}
where $z = q^+/P^\prime{}^+$ and we omit the term proportional to $\Gt$ since we are now conditioned to spot contributions it removes. 

First off, we know that contributions for $z$ outside of zero to one will require crossing of the final state vertex. Furthermore, in this situation no utilization of the bound state equation can remove the interaction already present. Thus to this order,  $0< z< 1$. Since now we deal with the final state 
vertex, a bit of notational adjustment is required. The interaction present in Eq. \eqref{jeff} can be written as
\begin{equation}
V(q,k+ \Delta) = \frac{- g^2}{2 P^\prime{}^+ (z - y)} \frac{1}{\qminus - \qminus_{v}(\kminus + \Delta^-)}, \notag
\end{equation}
and the imaginary part of the pole is $-\frac{\ie}{z - y}$. The remaining poles of the $\qminus$ integral are $\qminus_{a}$ which comes from $d(q)$ and has a form we are well familiar with, and $\qminus_{b}$ which comes from $d(P + \Delta - q)$. To avoid confusion, explicitly we see
\begin{equation}
\qminus_{b} = P^- + \Delta^- + \Big(q - (P + \Delta) \Big)^-_{\text{on}} - \frac{\ie}{2 P^\prime{}^+ (z - 1)}.
\end{equation}

In this form, the $\qminus$ integral can be evaluated just like many of the integrals above
\begin{equation}
\int d\qminus \frac{\gamma^*(z, \qperp - z \mathbf{\Delta}^\perp) V(q, k + \Delta)}{(\qminus - \qminus_{a})(\qminus - \qminus_{b})} \propto
\psi^*(z,\qperp - z \mathbf{\Delta}^\perp) \tilde{V}(\kminus, y, \kperp + \mathbf{\Delta}^\perp; z, \qperp).
\end{equation}
Hidden in $\tilde{V}$ are two cases depending on the sign of $z-y$, namely
\begin{equation}
\tilde{V}(\kminus, y, \kperp + \mathbf{\Delta}^\perp; z, \qperp) = \frac{g^2}{2 P^\prime{}^+ (y - z)} \Bigg( \frac{\theta(z - y)}{\kminus - k^{\prime -}_{v_{b}}} + \frac{\theta(y - z)}{\kminus - k^{\prime -}_{v_{a}}} \Bigg),
\end{equation}
where the primes are to keep clear the dependence on the final state momentum $P^\prime$ rather than the $P$'s heretofore. It is useful for what follows in section \ref{gpd} to spell out the primed poles
\begin{equation} 
	\begin{cases}
		k^{\prime -}_{v_{a}} = - \Delta^- + \qminus_{\text{on}} - \frac{\Big( \qperp - (\kperp + \mathbf{\Delta}^\perp) \Big)^2 + \mu^2}{2 [ \qplus - (\kplus + \Delta^+) ]} + \ie \frac{y}{z(z-y)}\\
		k^{\prime -}_{v_{b}} =  P^- + \Big( q - (P + \Delta) \Big)^-_{\text{on}} - \frac{\Big( \qperp - (\kperp + \mathbf{\Delta}^\perp) \Big)^2 + \mu^2}{2 [ \qplus - (\kplus + \Delta^+) ]} + \ie \frac{y - 1}{(z - 1) (z - y)}.
	\end{cases}
\end{equation}

This leaves us to evaluate the $\kminus$ integral. For $y>0$ and $y>z$, the poles of the integrand 
$\{\kminus_{a}, \kminus_{b}, \kminus_{c} k^{\prime -}_{v_{a}} \}$ all have negative imaginary parts except for 
$\kminus_{b}$. Taking the residue, produces the interaction $\kminus_{b} - k^{\prime -}_{v_{a}}$ which is half of the one-boson exchange potential 
for the final state. This contribution then will be removed by $\Gt$ when paired with the other exchange. 

Now consider the case $y>0$ and $y<z$ for which the poles are $a, b, c$ and $v^\prime_{b}$. In evaluating the integral, we find 
$- 2 \pi i \Big( \Res(\kminus_{a}) + \Res(\kminus_{c}) \Big)$. Neither of these residues corresponds individually to a time ordered contribution.  After algebraic manipulation into two new terms, however, one corresponds to the other boson exchange for the final state and is thus removed by $\Gt$ when paired with the term found above. The other term gives the contribution
 \begin{multline} \label{fs1}
 \frac{- 2 i P^+}{[2(2\pi)^3]^2} \int \frac{dx d\kperp \; \theta(x - \zeta)}{y x (1-x)} \; \frac{dz d\qperp \; \theta(z - y)}{z (1-z)} \; \frac{y - z}{g^2}
 \psi^*(z, \qperp - z \mathbf{\Delta}^\perp) \tilde{V}(\kminus_{c},y,\kperp+\mathbf{\Delta}^\perp;z,\qperp)\\
 \times \Big( 2 x - \zeta \Big) \tilde{V}(\kminus_{a},y,\kperp + \mathbf{\Delta}^\perp; z, \qperp)  \psi(x,\kperp),
 \end{multline}  
 which can be identified as diagram I of Figure \ref{ftri2}. 

 Finally there is one remaining case, $y<0$ (which automatically implies $z>y$). Only $\kminus_{a}$ has a negative imaginary part and hence we pick up $- 2 \pi i \Res(\kminus_{a})$. The final contribution at next to leading order is
 \begin{multline} \label{fs2}
 \frac{+ 2 i  P+}{[2 (2\pi)^3]^2} \int \frac{dx d\kperp \; \theta(\zeta - x)}{\sigma (1-\sigma)(1-x) \zeta} \frac{dz d\qperp \; \theta(z - y)}{z(1-z)} \psi^*(z, \qperp - z \mathbf{\Delta}^\perp) \\
 \times \dw(\sigma, \kperp + \sigma \mathbf{\Delta}^\perp | \Delta^2) \Big( 2 x - \zeta \Big) \tilde{V}(\kminus_{a}, y, \kperp + \mathbf{\Delta}^\perp; z, \qperp) \psi(x, \kperp),
 \end{multline}
 which corresponds to the time ordered diagram K in Figure \ref{fZZZ}. Figures \ref{ftri2} and \ref{fZZZ} then exhaust the contributions to the form factor at
 next to leading order. Thus in time ordered perturbation theory, we see there are no contributions from the non-wave function vertex---its presence as a reducible contribution was subtracted. We comment that current conservation is perturbatively improved for form factors at next to leading order over the Drell-Yan formula. Lastly if one is solely concerned with form factors, one can obtain simpler expressions by taking $\zeta \to 0$. 

 \section{Application to GPD's}\label{gpd}
Having worked through matrix elements of the electromagnetic current in a frame where $\Delta^+ \neq 0$, we can now make the connection to the generalized parton distributions (GPD's) for this model. These distributions, which in some sense are the natural interpolators between form factors and quark distribution functions, turn up in a variety of hard exclusive processes, e.g.  deeply virtual Compton scattering, wide-angle Compton scattering and the electroproduction of mesons \cite{Muller:1994fv}. The scattering amplitude for these processes factorizes into a convolution
of a hard part (calculable from perturbative QCD) and a soft part which the GPD's encode. Since light cone correlations are probed in these 
 hard processes, the soft physics has a simple interpretation and expression in terms of light front wave functions \cite{Brodsky:2001xy}.
In this section, we recast our results for form factors in section \ref{current} in the language of GPD's
and the light cone Fock space expansion. Additionally one can obtain these results directly from time ordered perturbation theory using two-body projection operators as explicated in Appendix C. 

 The GPD for our meson model is defined by a nondiagonal matrix element of bilocal field operators
 \begin{equation} \label{bilocal}
 F(x, \zeta, t) = \int \frac{dy^- e^{i x P^+ y^-}}{2 \pi (2 x - \zeta)}
 \langle \Psi_{P^\prime} | \; q(y^-)   i \overset{\leftrightarrow}\partial{}^+ q(0) \; | \Psi_{P} \rangle,
 \end{equation}
 where $q(x)$ denoted the quark field operator. Comparing to the current matrix element defined above \eqref{formula}, the definition of the GPD leads us immediately to the sum rule
 \begin{equation} \label{sumrule}
 \int \frac{2 x - \zeta}{2 - \zeta} F(x, \zeta, t) dx = F(t).
 \end{equation}
 Hence one can calculate these distributions from the integrand of the form factor. In this way, the light cone correlation defined in Eq. \eqref{bilocal} has a natural description in terms of light front time ordered perturbation theory, e.g. for $x>\zeta$ the relevant graphs contributing to the GPD are 
in Figure \ref{ftri} and \ref{ftri2}, and those for $x<\zeta$ are in Figure \ref{fZZZ}. Also note, the $\zeta$ dependence disappears after the $x$ integration present in the sum rule due to Lorentz invariance. We were aware of this in section \ref{current} when we opted not to choose a frame in which $\Delta^+ = 0$ for the form factor when we knew at the end of the day $F(t)$ would be a function of $t = \Delta^2$ alone.

\subsection{Continuity}
Conversion of the contributions to the form factor above into GPD's is straightforward using Eq. \eqref{sumrule} and \eqref{formfactor}; we merely 
remove $- i P^+ (2 x - \zeta)$ and $\int dx$ from Eqs. \eqref{bt1}, \eqref{bt2}, \eqref{is1}, \eqref{fs1}, \eqref{fs2} 
and $\int dx \frac{2x - \zeta}{2 - \zeta}$ from equation 
\eqref{FFLO}. In order for the deeply virtual Compton scattering amplitude to factorize into hard and soft pieces (at leading twist), 
the GPD's $F(x,\zeta,t)$ must be continuous at $x = \zeta$. Continuity at the crossover is an even more pressing matter since all current 
experimental efforts to measure such functions are limited to the crossover \cite{Diehl:1997bu}. The leading order expressions are continuous. This 
is easy to see since the contribution for $x < \zeta$ is identically zero. The valence contribution is a convolution of wave functions
one of which is $\psi^*(y,\ldots)$ which is probed at the end point since $y \to 0$. From the bound state equation Eq. \eqref{wavefunction}, 
we see the two-body wave function vanishes quadratically at the end points. Taking into account the overall weighting factor of $y^{-1}$, the 
valence piece vanishes linearly at the crossover. At leading order then, continuity is maintained at the crossover, while the derivative
is discontinuous. Working only in the valence sector, valence quark models will never be of any use to experiment since the value at the 
crossover requires one wave function to be at an end point. In the three-body bound state problem (e.g. the nucleon), 
the wave function again necessarily vanishes; however, 
the valence GPD will vanish only if the three-body interaction vanishes at the end points (which is physically reasonable). 

Let us now check the next to leading order contributions to the GPD for continuity. First we shall deal with the term stemming from iterating 
the Bethe-Salpeter equation for the initial state \eqref{is1}. Since there is no Z graph generated from iterating the initial state, we expect this 
contribution to vanish; after all, there is no related piece for $ x < \zeta$ to match up with. Looking at the expression, we see again 
$\psi^*(y,\ldots)/y$ which vanishes linearly as $x \to \zeta$. Moreover there are the interaction terms $\theta(w - x) 
\tilde{V}(\kminus_{a},x,\kperp; w,\pperp)$, which is finite as $x \to \zeta$, and 
\begin{equation}
\theta (w - x) \tilde{V}(\kminus_{c}, x, \kperp; w, \pperp) \overset{x\to \zeta}{=} \frac{-g^2}{w - \zeta} 
\; \frac{x - \zeta}{(\kperp + \mathbf{\Delta}^\perp)^2 + m^2}, 
\end{equation}
which vanishes at the crossover. Thus not only does the initial state iteration term vanish at the crossover, its derivative does so as well.

Now we investigate the Born terms \eqref{bt1} and \eqref{bt2} at the crossover. Approaching $\zeta$ from above \eqref{bt1}, we encounter
two interactions $\tilde{V}(\kminus_{a}, \ldots)$ which is finite as above and $\tilde{V}(\kminus_c, \ldots)$ which vanishes linearly (enough
to cancel the weight $(x - \zeta)^{-1}$. Thus we are left with the finite contribution to the GPD at the crossover from the Born term
\begin{multline}\label{bcross}
F(\zeta, \zeta, t)^{\text{Born}} = \frac{2}{[2(2\pi)^3]^2} \int \frac{d\kperp}{\zeta} \;  \frac{dw d\pperp \theta(w - \zeta)}{v w (1-w)}
\\ \times \psi^*(v, \pperp + (1-v) \mathbf{\Delta}^\perp) \frac{1}{(\kperp + \mathbf{\Delta}^\perp)^2 + m^2} 
\tilde{V}\Big(\frac{\kperp{}^2 + m^2}{2P^+ \zeta},\zeta;w,\pperp \Big) \psi(w, \pperp). 
\end{multline} 

On the other hand, approaching the crossover from below \eqref{bt2} we have to deal with singularities as $\sigma = x/ \zeta \to 1$. Writing out
the propagator for the quark anti-quark pair heading off to annihilation, we see
\begin{equation} \label{qqbarprop}
\dw(\sigma, \kperp + \sigma \mathbf{\Delta}^\perp | \Delta^2 ) \to - \frac{1 - \sigma}{(\kperp + \mathbf{\Delta}^\perp)^2 + m^2}.
\end{equation}
The linear vanishing of this term is enough to cancel the singular weight $(1 - \sigma)^{-1}$. Taking the limit $x \to \zeta$ then produces
equation \eqref{bcross} and thus the Born terms are continuous. 

Lastly we must see how the final state iteration terms match up at the crossover. Having spelled out the Born terms, the final state
terms follow simply once we note $\tilde{V}(\kminus_{a}, y, \kperp + \mathbf{\Delta}^\perp; z, \qperp)$ is finite as $y \to 0$ and
\begin{equation}
\theta(z - y) \tilde{V}(\kminus_{c}, y, \kperp + \mathbf{\Delta}^\perp; z, \qperp) \to \frac{-g^2 y}{z} \frac{1}{(\kperp + \mathbf{\Delta}^\perp)^2 + m^2},
\end{equation}  
which vanishes as $x$ goes to $\zeta$. Thus at the crossover we have the contribution
\begin{multline}\label{fcross}
F(\zeta, \zeta, t)^{\text{final}} = \frac{2}{[2(2\pi)^3]^2} \int \frac{d\kperp}{\zeta(1-\zeta)} \; \frac{dz d\qperp}{z(1-z)} 
\\ \times \psi^*(z, \qperp - z \mathbf{\Delta}^\perp) \frac{1}{(\kperp + \mathbf{\Delta}^\perp)^2 + m^2} 
\tilde{V}\Big( \frac{\kperp^2 + m^2}{2 P^+ \zeta}, 0, \kperp + \mathbf{\Delta}^\perp; z, \qperp \Big) \psi(z, \qperp).
\end{multline} 
Approaching $\zeta$ from below, we again utilize equation \eqref{qqbarprop} in taking the limit of \eqref{fs2}. The result is \eqref{fcross}
and hence we have demonstrated the GPD's continuity to first order in the weak coupling, i.e.
\begin{equation}
F(\zeta,\zeta,t) = F(\zeta,\zeta,t)^{\text{Born}} +   F(\zeta,\zeta,t)^{\text{final}}
\end{equation}
no matter how we approach $x = \zeta$. 

\subsection{Fock space representation}
We now proceed to rewrite the GDA's in terms of Fock component overlaps. In the diagonal overlap region $x > \zeta$ this will be a mere rewriting 
of our results, while there is a subtlety for the nondiagonal overlaps. First let us rewrite the term to zeroth order in the weak coupling \eqref{FFLO}.
Define the two-body Fock component as
\begin{equation} \label{twofock}
\psi_{2}(x_{1},\kperp_{1}, x_{2}, \kperp_{2}) = \frac{1}{\sqrt{ x_{1} x_{2} }} \psi(x_{1}, \kperp_{\text{rel}}), 
\end{equation}
noting that the relative transverse momentum can be defined as $\kperp_{\text{rel}} = x_{2} \mathbf{k}^\perp_{1} - x_{1} \mathbf{k}^\perp_{2}$. In terms of Eq. \eqref{twofock},
the GPD appears as 
\begin{equation} \label{222}
F(x, \zeta, t)^{\text{LO}} = \frac{\theta(x - \zeta)}{\sqrt{1-\zeta}} \int [dx]_{2} [d\kperp]_{2} \sum_{j = 1,2}\frac{ \delta(x - x_{j})}{\sqrt{x^{\prime}_{j} x_{j}}}  \psi_{2}^*(x^{\prime}_{i}, \mathbf{k}^\prime_{i}{}^\perp) 
\psi_{2}(x_{i},\mathbf{k}_{i}^\perp),  
\end{equation}
where the primed variables are given by
\begin{equation} \label{primed}
\begin{cases}
x^\prime_{i} = \frac{x_{i}}{1-\zeta}\\
\mathbf{k}^\prime_{i}{}^\perp  = \mathbf{k}_{i}^\perp - x^\prime_{i} \mathbf{\Delta}^\perp, \; \text{for} \; i \neq j
\end{cases}
\begin{cases}
x^\prime_{j} = \frac{x_{j} - \zeta}{1 - \zeta}\\
\mathbf{k}^\prime_{j}{}^\perp = \mathbf{k}_{j}^\perp + (1 - x^\prime_{j}) \mathbf{\Delta}^\perp
\end{cases}
\end{equation}
and the integration measure is given by
\begin{align}
[dx]_{N} & = \prod_{i = 1}^{N} dx_{i} \; \delta \Big( 1 - \sum_{i = 1}^{N} x_{i} \Big)\\
[d\kperp]_{N} & = \frac{1}{[2(2\pi)^3]^{N-1}} \prod_{i=1}^N d\mathbf{k}^\perp_{i} \; \delta \Big( \sum_{i=1}^N \mathbf{k}^\perp_{i} \Big).
\end{align}
Notice the sum over transverse momenta in the delta function is zero since our initial meson has $\mathbf{P}^\perp = 0$.
The sum over $j$ in Eq. \eqref{222} produces the overall factor of two for our case of equally massive (equally charged) constituents. 

To cast the next to leading order expressions for $x > \zeta$ in terms of diagonal Fock space overlaps, we need to write out the three-body Fock component.
Looking at the diagrams in Figure \ref{ftri2}, it is constructed from the two-body wave function
\begin{multline} \label{threefock}
\psi_{3} (x_{i},\mathbf{k}_{i}^\perp) = \frac{2 (2\pi)^3}{\sqrt{x_{1} x_{3}}} \sqrt{\frac{x_{2}}{g^2}} 
\int [dy]_{2} [d\pperp]_{2} \Bigg[ \theta(y_{1} - x_{1}) x_{3} \delta(y_{2} - x_{3}) 
\delta(\mathbf{p}^\perp_{2} - \mathbf{k}^\perp_{3}) \tilde{V}(k^-_{1}{}_{\text{on}},x_{1},\mathbf{k}^\perp_{1};x_{3},\mathbf{k}^\perp_{3}) \\
+ \theta(y_2 - x_3) x_1 \delta(y_{1} - x_{1}) \delta(\mathbf{p}^\perp_{1} - \mathbf{k}^\perp_{1}) \tilde{V}(k^-_{3}{}_{\text{on}},x_{3},\mathbf{k}^\perp_{3};x_{1},\mathbf{k}^\perp_{1}) \Bigg] \frac{\psi_{2}(y_{j},\mathbf{p}_{j}^\perp)}{\sqrt{y_{1} y_{2}}},
\end{multline}
where $i$ runs from one to three and the label $j$, which stems from the integration measure, runs from one to two. The awkward looking factor of 
$\sqrt{x_2/g^2}$ arises from our definition of $\tilde{V}$. It serves to remove a factor of $g$ since only one interaction is needed to place us in the 
three-body subspace and when combined with the prefactor in $\tilde{V}$ produces the factor $1/\sqrt{x_{2}}$ which then symmetrically appears
with the overall constant in $\psi_{3}$.  We discuss how to obtain this three-body wave function directly from time ordered perturbation theory in 
Appendix C. Using $\psi_3$ in Eq. \eqref{threefock}, the terms in the GPD at first 
order in the weak coupling can then be written compactly as
\begin{equation} \label{323}
F(x,\zeta,t)^{\text{NLO}} = \frac{\theta(x - \zeta)}{1 - \zeta} \int [dx]_3 [d\kperp]_{3} \sum_{j = 1,3} \frac{\delta( x - x_{j})}{\sqrt{x^\prime_{j} x_{j}}} 
\psi_{3}^*(x_{i}^\prime,\mathbf{k}_{i}^\prime{}^\perp) \psi_{3}(x_{i},\mathbf{k}^\perp_i).
\end{equation}
One can verify that the diagrams in Figure \ref{ftri2} are generated by \eqref{323}. Note well that our definition of $z$
as a momentum fraction with respect to the final plus momentum $P^\prime{}^+$ already includes the factor of $(1-\zeta)^{-1}$ with respect 
to $P^+$. Additionally there is a fourth diagram generated by Eq. \eqref{323} which does not appear in the figure. This missing diagram is characterized
by the spectator quark's one-loop self interaction. Thus in performing calculations in perturbation theory, we should also use
the one-loop result for the renormalized self interactions $f(k^2)$ appearing in the spectator's propagator. As far as our development is concerned, 
we imagined this contribution tacked onto Figure \ref{ftri}, while it appears now as part of the three-to-three Fock component overlaps. The absence of 
this diagram does not affect continuity at the crossover. The missing diagram vanishes at $x = \zeta$ since the final
state wave function is $\psi^*(y,\ldots)$. 

Now we must come to terms with the nondiagonal overlap region, $x < \zeta$. At first order in the weak coupling, the diagrams of Figure \ref{fZZZ}
correspond to four-to-two Fock component overlaps. We have been cavalier about time ordering, however. The expressions we derived in 
Eqs. \ref{bt2} and \ref{fs2} do not actually correspond to time ordered graphs. Both terms contain a product of time ordered propagators:
one for the two quarks leading to the final state vertex and another for the quark-anti-quark pair heading off to annihilation. But for an interpretation
in terms of a four-body wave function, all four particles must propagate at the same time. This is a subtle issue as a graph containing the product 
of two independently time ordered pieces (where one leads to a bound state vertex) corresponds to a sum of infinitely many time ordered graphs. Writing out the terms that concern us, we can 
manipulate as follows
\begin{align} 
- (2 P^+)^2 \zeta (1-\zeta) \dw(\sigma, \kperp + \sigma \mathbf{\Delta}^\perp |t) \dw(z, \qperp - z \mathbf{\Delta}^\perp|M^2) & = 
\frac{1}{\kminus_{c} - \kminus_{a}} \; \frac{1}{\qminus_{b} - \qminus_{a}} \\
	& =  \frac{1}{\qminus_{b} - \qminus_{a} + \kminus_{c} - \kminus_{a}} \; \Bigg( 
\frac{1}{\qminus_{b} - \qminus_{a}} + \frac{1}{\kminus_{c} - \kminus_{a}} \Bigg).   \label{trickery}
\end{align}
In this form, we have produced the correct energy denominator for the instant of light front time where four particles are propagating. 
Multiplying this denominator by the three-body wave function yields the four-body wave function (up to constants).
This is only the part of the four-body wave function relevant for GPD's (there are additional pieces for two-quark, two-boson states, see Appendix C).  
In the resulting sum \eqref{trickery}, the first term will produce the two-body wave function for the final state and we will have a
genuine four-to-two overlap. We do not write this out explicitly.

The second term in Eq. \eqref{trickery}, however, contains again the propagator for the pair heading to annihilation. Using the light front 
Bethe-Salpeter equation for the vertex (which contains infinitely many times) we can introduce a factor of the time ordered interaction. 
The resulting product of independent time orderings can again be manipulated as in Eq. \eqref{trickery}. The result produces another overall four-body
denominator which contributes to the four-body Fock component of the initial state. Since we iterated the interaction, however, this new contribution
is no longer at leading order and can be neglected. Thus the second term in \eqref{trickery} does not contribute at this order.

Having manipulated the GPD's into nondiagonal overlaps for $x < \zeta$, we must wonder if continuity at the crossover is still maintained. In the limit 
$x \to \zeta$ the light front energy of the struck quark goes to infinity (since it has vanishing plus momentum). Consequently $\kminus_{c}$, 
which contains this on shell energy, also is infinite and dominates the four-body energy denominator. This is identical to the reasoning
above in Eqs. \ref{bt2} and \ref{fs2} where instead of the four-body denominator, we have $\dw(\sigma, \kperp + \sigma \mathbf{\Delta}^\perp |t)$ which 
is dominated by $\kminus_{c}$ at the crossover. In both cases we arrive at the expressions found above for the crossover \eqref{bcross} and \eqref{fcross}.

Having cast our expressions for generalized parton distributions in terms of the Fock components generated to first order in the weak coupling, we
can enlarge our understanding of the sum rule and continuity at the crossover. Both must deal with the relation between higher Fock components. The way 
$\zeta$ dependence disappears from \eqref{sumrule} mandates a relation between the diagonal and nondiagonal Fock component overlaps that make up the 
GPD. The relation between Fock components must follow from the field theoretic equations of motion. The continuity condition itself is a special
case of the relation between Fock components, specifically at the end points. Above we have seen our expressions are continuous (and non vanishing)
at the crossover and explicitly that the three- and four-body components match at the end point (where $x - \zeta = 0$). This weak binding model for behavior 
at the crossover is a simple example of the relations which exist between Fock components at the end points \cite{Antonuccio:1997tw}. More general 
relations must be permitted from the equations of motion to guarantee Lorentz covariance (e.g. in the structure of the Mellin moments of the GPD's, of which 
the sum rule is a special case).

 \section{Application to GDA's}\label{gda}
 In section \ref{current} we worked through spacelike form factors in the reduced Bethe-Salpeter formalism, then in section \ref{gpd} we connected these
 results to GPD's. Below we tackle timelike form factors to demonstrate the versatility of this approach and make the connection to the generalized
 distribution amplitudes (GDA's) for this model. Analogous to GPD's, GDA's encode the soft physics of two-meson production and can thus be thought of 
 as  crossed versions of the GPD's. The GDA's enter in convolutions for various two-meson production amplitudes \cite{Diehl:1998dk}. These distribution functions as well as timelike form factors are a theoretical challenge for light front dynamics, since there is no direct decomposition in terms of meson Fock components alone. Furthermore, we shall see the leading order expressions are nonvalence contributions (which necessarily excludes a description in terms of most 
constituent quark models). 

 \begin{figure}
 \begin{center}
	\epsfig{file=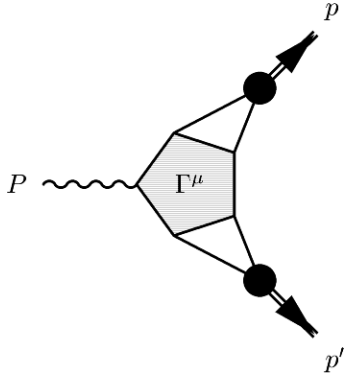}
\caption{The triangle diagram for the timelike, pion form factor}
\label{ftimetri}
\end{center}
\end{figure}

The timelike form factor $F(s)$ for our model meson is defined by (see Figure \ref{ftimetri})
\begin{equation}
\langle \Psi_{p} \; \Psi_{p^\prime} | \; \Gamma^\mu \; | 0 \rangle = - i (p - p^\prime)^\mu F(s),
\end{equation}
where $s = (p + \pp)^2$ is the center of mass energy squared. Now define $P^\mu = p^\mu + \pp{}^\mu$ and $\zeta = \pplus / \Pplus$. We can work out the kinematics of this reaction in a frame where $\mathbf{P}^\perp = 0$ 
\begin{align}
P^- & = \frac{s}{2 P^+} \notag \\
p^- & = \frac{(1 - \zeta) s}{2 \Pplus} \notag \\
\mathbf{p}^\perp{}^2 & = s (1 - \zeta) \zeta - M^2,
\end{align}
where $M$ is the meson mass. 

Similar to GPD's, the GDA for our model has a definition in terms of a nondiagonal matrix element of bilocal field operators
\begin{equation}
\GDA = \int \frac{dx^- e^{i z \Pplus x^-}}{2 \pi (2 z - 1)} \langle \Psi_{p} \; \Psi_{\pp} | \; q(x^-) i \
\overset{\leftrightarrow}\partial{}^+ q(0) \; | 0 \rangle.
\end{equation}
Such a definition of the GDA leads directly to the sum rule for the time like form factor
\begin{equation} \label{sumtime}
\int \frac{2 z - 1}{2 \zeta - 1} \GDA dz = F(s),
\end{equation}
and hence a means to calculate $\Phi$ from the integrand of the timelike form factor. 

Taking the appropriate residues of the five-point function, we arrive at Fig. \ref{ftimetri} for the timelike form factor. Keeping only the leading order piece of the electromagnetic vertex $\Gamma^\mu$, we have
\begin{equation}
\GDA = 2 \Pplus  \int \frac{d\kminus d\kperp}{(2\pi)^4} \gamma^*\Big( \frac{z}{\zeta}, \kperp - \frac{z}{\zeta} \pperp \Big| M^2 \Big) G(k,p) d^{-1}(k-p) G(k-p, \pp) \gamma^*\Big( \frac{z-\zeta}{1 - \zeta}, \kperp - \frac{1-z}{1-\zeta} \pperp    \Big| M^2 \Big), 
\end{equation}
where we have made use of $z = \kplus/ \Pplus $. Since this term carries no overall factor of the coupling, we must have only wave function vertices (knowing well that any apparent non-wave function vertices will be subtracted in the complete next order expression). This requirement translates to $0< \frac{z}{\zeta} < 1$ and $0< \frac{z - \zeta}{1 - \zeta} < 1$, and hence we do not pick up a contribution at zeroth order in the coupling. 

To work at first order, we pick up three terms analogous to those in section \ref{current}. We denote these as $\delta J^\mu_{\gamma}, \delta J^\mu_{p}$ and
$\delta J^\mu_{\pp}$. The first Born term for the three-point electromagnetic vertex $\delta J^\mu_{\gamma}$ is quite simple. Any diagrams contributing to this term must contain an overall power of the weak coupling since the boson can never be reduced into either final state vertex. Thus we must restrict the plus momentum fractions of the two final state vertices to be between zero and one. For the same reason as the zeroth order result, momentum conservation in the diagram shows this restriction cannot be met.  This leaves us to consider only diagrams that arise from iteration of the Bethe-Salpeter equation of either final state meson.

Considering first the term $\delta J^\mu_{p}$, we have the contribution to the GDA
\begin{equation} \label{blah}
\Phi_{p}(z,\zeta,s)  =  2 i \Pplus \int \frac{d\kminus d\kperp}{(2\pi)^4} \; \frac{d^4q}{(2\pi)^4}
\gamma^*(w, \qperp - w \pperp) G(q,p) V(q.k) G(k,p) G(k - p, \pp) d^{-1}(k-p)
\gamma^*(y, \kperp - (1-y) \pperp), 
\end{equation}
where we have chosen to abbreviate $w = \qplus/ \pplus$, $ y = \frac{z - \zeta}{1 - \zeta}$. We have customarily omitted the subtracted term containing $\Gt$ since we know it will only remove the non-wave function vertices present in Eq. \ref{blah} which have the same for as those encountered at zeroth order. Thus we can entirely omit any non-wave function vertices, knowing well they will be canceled. Thus we have $0<w<1$ and $\zeta < z < 1$, which will be of incredible aid in evaluating the integrals. 

\begin{figure}
\begin{center}
\epsfig{file=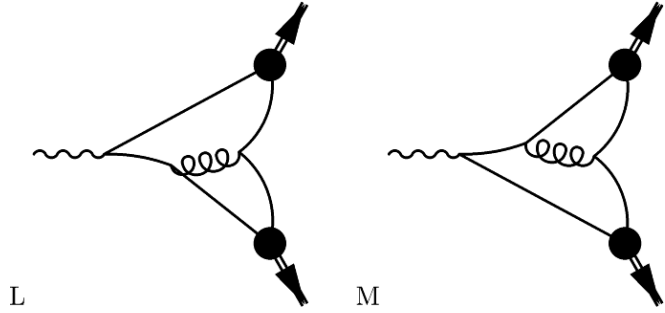}
\caption{Leading order diagrams for the timelike, pion form factor.}
\label{fgda}
\end{center}
\end{figure}

With these restrictions, the $\qminus$ integral can be performed similar to equation \eqref{firsto} and the subsequent $\kminus$ integration resembles that in section \ref{current} but here all the imaginary parts' signs are fixed. 
Thus we get one contribution $\Phi_{p}$ to the GDA from $J^\mu_{p}$
\begin{multline} \label{jp}
\Phi_{p}(z,\zeta,s)  = \frac{2}{[2 (2\pi)^3]^2} \int \frac{d\kperp \; \theta(z - \zeta)}{y z (1-z)} \; \frac{dw d\qperp}{w(1-w)} \psi^*(w, \qperp - w \pperp) \\ \times \tilde{V}\Big(\kminus_{b},\frac{z}{\zeta}, \kperp; w, \qperp \Big) \dw(z, \kperp |s) \psi^*(y, \kperp - (1-y) \pperp),  
\end{multline}
where for clarity we spell out the form of the interaction
\begin{equation} 
\frac{g^2 \theta(z - \zeta)}{2 \pplus \Big( \frac{z}{\zeta} - w  \Big)} \tilde{V} \Big(\kminus_{b} , \frac{z}{\zeta}, \kperp; w, \qperp\Big)^{-1}  = 
P^- - q^-_{\text{on}} - \frac{(\kperp - \qperp)^2 + \mu^2}{2 (\kplus - \qplus)} - (P - k)^-_{\text{on}}.
\end{equation}
As far as contributions to the timelike form factor are concerned, we can interpret Eq. \eqref{jp} as the time ordered diagram M of Figure \ref{fgda}. 

At first order in the weak coupling, we have one term remaining to consider $\delta J^\mu_{\pp}$. Omitting the subtraction of $\Gt$, we have

\begin{multline}
\Phi_{p^\prime}(z,\zeta,s) = 2 i \Pplus \int \frac{d\kminus d\kperp}{(2\pi)^4} \; \frac{d^4 q}{(2 \pi)^4} \gamma^*\Big(\frac{z}{\zeta}, \kperp - \frac{z}{\zeta} \pperp \Big| M^2 \Big) G(k,p) d^{-1}(p-k) \\ 
\times G(P - k , \pp) V(P-k, q) G(q, \pp) \gamma^*(w, \qperp + w \pperp|M^2). 
\end{multline}
Since we are removing the crossed interactions by hand, we restrict both 
$0 < w < 1$ and $0 < z < \zeta$. The signs of all poles' imaginary parts
are now fixed and the integration proceeds just as many above. We thus arrive at the final contribution to the GDA
\begin{multline} \label{jpp}
\Phi_{p^\prime}(z,\zeta,s) = \frac{2}{[2 (2\pi)^3]^2} \int \frac{d\kperp \; \theta(\zeta - z)}{\sigma (1 - \sigma) \zeta} \; \frac{dw d\qperp}{w(1-w)} \psi^*(\sigma, \kperp - \sigma \pperp) 
\\ \times \dw(z, \kperp | s) \tilde{V}(\kminus_{a}, 1 - z, -\kperp; w, \qperp) \psi^*(w, \qperp + w \pperp), 
\end{multline} 
where we have defined $\sigma = z / \zeta$ and the interaction is
\begin{equation}
 \frac{g^2 \theta(\zeta - w) }{2 \pp{}^+ \Big( \frac{1 - z}{1 - \zeta} - w \Big)} \tilde{V}(\kminus_{a}, 1 - z, -\kperp; w, \qperp)^{-1} = 
P^- - \kminus_{\text{on}} - [(P - k) - q]^-_{\text{on}} - q^-_{\text{on}} .
\end{equation}
In this form we recognize this contribution as diagram L of Figure \ref{fgda}. 

Having found the leading non-vanishing contribution to the GDA namely $\Phi = \Phi_{p} + \Phi_{p^\prime}$, we observe that the higher Fock components derived in section \ref{gpd} (as well as in Appendix C) do not fit naturally into \eqref{jp} or \eqref{jpp}. One needs a Fock space expansion for the photon wave function in order to have an expression for the GDA in terms of various Fock component overlaps. With the expressions derived for the GDA we can use Eq. \eqref{sumtime} to obtain the timelike form factor. 

\section{Conclusion} \label{conclusion}
We have motivated the light front reduction of the Bethe-Salpeter equation which yields a valence wave function calculated 
from a light front, time ordered kernel. Intuitively this is performed by iterating the covariant equation any number of times
followed by a single instantaneous approximation that allows for all of the light front energy integrals to be performed. Formally it is achieved
by using an auxiliary Green's function $\Gt$ defined in Eq. \eqref{gt}. The reduction scheme for the Bethe-Salpeter equation 
is then applied to matrix elements of the electromagnetic current. Writing out the gauge invariant current completely Eq. \ref{emvertex}, we were 
able to investigate a bound state's interaction with an electromagnetic probe systematically in perturbation theory. We illustrated applications
to form factors, generalized parton distributions and generalized distribution amplitudes.

To carry out our analysis, we were forced to adopt the ladder approximation for the interaction \eqref{ladder}.  With this model, we saw
how further iteration of the covariant Bethe-Salpeter equation leads to a better approximation to the energy poles of the vertex function, 
\emph{cf} Eq. \eqref{V2} for the effective two-body interaction, which has a second order term stemming from the four-dimensional 
nature of the kernel. Alternately, we can view the energy poles of the vertex as giving rise to contributions from the nonvalence Fock space, 
i.e. the additional poles which allow for the second order correction to the interaction also have a higher number of quanta propagating
during intermediate light cone times.  To make this connection more transparent, we saw how the energy poles generated the higher
Fock space contributions to the form factor in section \ref{current} (written out explicitly in \ref{gpd}). Additionally in Appendix A, 
we saw in Eq. \ref{nonN} how corrections to the normalization stemming from poles of the vertex could be interpreted as a nonvalence 
probability \eqref{quspace}.

In obtaining matrix elements of the electromagnetic current, we were confronted with Z graphs containing non-wave function vertices, 
see Eq. \eqref{evil}, for example. Within time ordered perturbation theory, where there are higher Fock components contained in
the Bethe-Salpeter wave function, these non-wave function vertices are removed when one works completely to any given order in perturbation 
theory. We contrasted this with the case of an instantaneous interaction (or most often encountered, contact term approximation for 
the Bethe-Salpeter vertex) in Appendix B. In this special case, one can successfully appeal to crossing in order to interpret non-wave 
function vertices; however, in this approach, there are no true higher Fock components. At first mention, it would seem
that one could use these types of models to get some glimpse of nonvalence physics. This might be the case for timelike processes, 
but for GPD's the distributions will always vanish at the crossover which is useless for comparing with experiment. Lastly one could use an instantaneous
reduction scheme to reduce the Bethe-Salpeter equation and still appeal to crossing to have non-zero, non-wave function vertices. 
In this situation, one is sacrificing accuracy in determining the light front wave function, since the potential is not 
calculated in time ordered perturbation theory. Moreover, the higher Fock pieces then become entangled in this instantaneous reduction, whereas
one believes that corrections to the form factor from higher Fock components are systematically smaller for higher particle content. From the
above analysis, we expect cancelations of the instantaneous terms in favor of higher Fock states. 

Having worked within time ordered perturbation theory, however, our expansion of the form factor (and generalized parton distributions)
at leading and next to leading order was economical. In this form, we were able to demonstrate that our GPD is continuous at the crossover
(while the first derivative is discontinuous). We also unraveled the generation of higher Fock states from the two-body sector using our
explicit expressions for GPD's. The diagonal overlaps could be written down directly, whereas to uncover the nondiagonal overlaps an infinite
sum of time orderings had to be dealt with using efficacious algebra. The expressions for higher Fock contributions agreed with those calculated from 
old-fashioned perturbation theory in Appendix C.  Moreover, we saw in section \ref{gpd}, how the higher Fock states lead to the non-vanishing
at the crossover and are hence essential for any phenomenological modeling of these distributions. Lastly we investigated timelike form factors and 
the related generalized distribution amplitudes in this model. Expressions for these were systematically evaluated in perturbation theory
which is in contrast to the non-existent Fock space expansion for these types of processes. 

With the formalism explored here, one could use phenomenological Lagrangian based models to explore both 
generalized parton distributions and generalized distribution amplitudes within the light front framework.
Of particular interest would be the breaking of Lorentz invariance (in particular, rotational invariance, see \cite{Cooke:1999yi})
characteristic of Hamiltonian theories. This breaking would manifest itself as a violation of the sum rules 
(and polynomiality conditions). Such an investigation is interesting not only for testing phenomenological models, 
but also for anticipating problems for any approximate non-perturbative solutions for the light cone Fock states. 
Nonetheless more model studies are warranted before truly realistic calculations can be pursued.

\begin{center}
{\bf Acknowledgment}
\end{center}
This work was funded by the U.~S.~Department of Energy, grant: 
DE-FG$03-97$ER$41014$.  

\section*{Appendix A: Normalization}
In section \ref{formal} the relation between the four dimensional Bethe-Salpeter wave function and the reduced light front wave function is presented. We did not, however, address the normalization of either. Given the relation between reduced and covariant quantities, we can use the Bethe-Salpeter wave function's 
normalization condition to derive the reduced version. First let us review the covariant wave function's normalization \cite{Itzykson:rh}.

The four-point function defined in Eq. \eqref{LSg4} has the behavior
\begin{equation}
G^{(4)}(R) =  - i \frac{|\Psi_{R} \rangle \langle \Psi_{R} |}{R^2 - M^2 + \ie} + \; \text{finite},
\end{equation} 
near the bound state pole. In what follows, we remove the total momentum labels on operators since they are all identical, $R$.
Since the four-point function satisfies the equation
\begin{equation}
G^{(4)} = G^{(4)} \Big( G^{-1} - V \Big) G^{(4)},
\end{equation}
the Bethe-Salpeter amplitude must satisfy
\begin{equation}
1 = \lim_{R^2 \to M^2} - i  \frac{\langle \Psi_{R}| \Big( G^{-1} - V \Big) |\Psi_{R}\rangle}{R^2 - M^2},
\end{equation}
which is necessarily finite since $|\Psi_{R}\rangle$ satisfies the bound state
equation: $|\Psi_{R}\rangle = G V |\Psi_{R}\rangle$. Application of l'H\^opital
yields the covariant normalization condition
\begin{equation} \label{normcov}
2 i R^\mu =  \langle \Psi_{R} | \frac{\partial}{\partial R_{\mu}} \Big( G^{-1} - V  \Big) |\Psi_{R} \rangle,
\end{equation}
with evaluation at $R^2 = M^2$ understood. 

The normalization condition for the reduced wave function is then deduced by using the conversion \eqref{324} in Eq. \eqref{normcov}. Hence taking the plus component
\begin{equation} \label{normlf}
2 i R^+ =  \langle \gamma_{R} | \; \Big| G \Bigg( 1 + W ( G - \Gt ) \Bigg) \Bigg( \frac{\partial}{\partial R^-} \Big[ G^{-1} - V  \Big] \Bigg) \Bigg( 1 + (G - \Gt ) W \Bigg) G \Big| \; |\gamma_{R}\rangle.
\end{equation}
The complicated normalization condition is indicative of the higher Fock space components contained in the Bethe-Salpeter wave function which must now be generated from $W$ when we deal exclusively with the reduced wave function. To see this explicitly, we again work in our perturbative model.

Let us start with the leading order contribution to the reduced wave function's normalization. 
\begin{equation} \label{normLO}
\frac{-i}{2 R^+}\langle \gamma_{R}  | \; \Big| G \Big( \frac{\partial}{\partial R^-} G^{-1} \Big)  G \Big| \; | \gamma_{R} \rangle = 1.
\end{equation}
To perform the integration, we note
\begin{equation}
\frac{\partial}{\partial R^-} G^{-1}(k,R) = - 2 i R^+ d^{-1}(k) (1 - x), 
\end{equation}
where we have customarily chosen $x = \kplus/R^+$. Evaluation of the integral in equation \eqref{normLO} is now standard given the two poles $\kminus_{a}$ and $\kminus_{b}$. Closing the contour in the lower-half plane for $x \in (0,1)$, we pick up the (simple) residue at $\kminus_{a}$ avoiding the double pole at $\kminus_{b}$. Sorting out the overall factors gives
\begin{equation} \label{2to2}
N^{\text{LO}} \equiv \int \frac{dx d\kperp}{2 (2\pi)^3 x (1-x)} \psi^*(x,\kperp) \psi(x,\kperp) = 1,
\end{equation}
a simple overlap of the two-body wave function. 

To analyze the normalization to first order in the weak coupling, we note that our covariant interaction is a function of the momentum difference alone, i.e.
$\partial V / \partial R^\mu  = 0$. As in section \ref{current}, we substitute $W \approx V$ to first order. Expanding equation \eqref{normlf} to first order we find 
\begin{equation} 
N^{\text{LO}} + \delta N  + \ldots = 1,
\end{equation}
where $N^{\text{LO}}$ is the result \eqref{2to2} and the first order correction is
\begin{equation} \label{deltaN}
\delta N = \frac{-i}{2 R^+} \langle \gamma_{R} | \; \Big| G \Big( \frac{\partial}{\partial R^-} G^{-1} \Big) \Big( G - \Gt \Big) V G + 
G V \Big( G - \Gt \Big) \Big( \frac{\partial}{\partial R^-} G^{-1} \Big) G \Big| \; | \gamma_{R} \rangle,
\end{equation}
where by Eq. \eqref{gt}, the presence of $\Gt$ merely subtracts the leading order result $N^{\text{LO}}$. Considering just the first term in the above equation (and omitting the subtraction $\Gt$), we have
\begin{equation}
- i \int \frac{d^4 k}{(2\pi)^4} \; \frac{d^4p}{(2\pi)^4} (1-x) \gamma^* (x,\kperp|M^2) d(k) d(R-k)^2 V(k,p) d(p) d(R-p) \gamma (y,\pperp|M^2).
\end{equation}
The $\pminus$ integral is identical to that considered in deriving the bound state equation to leading order in section \ref{intuitive}. Thus using \eqref{vtilde} we have
\begin{equation}
- \int \frac{d^4 k}{(2\pi)^4} (1-x) \gamma^*(x,\kperp|M^2) d(k) d(R-k)^2 \int \frac{dy d\pperp}{2 (2\pi)^3 y (1-y)} \tilde{V}(\kminus, x, \kperp; y, \pperp) 
(-i) \psi(y, \pperp).
\end{equation}
The poles confronting us are again the same as those for the $\kminus$ integral of the first order bound state equation (the only difference of course is that $\kminus_{b}$ is now a double pole). When $y>x$, we pick up the residue at $\kminus_{a}$ similar to before, which not surprisingly will be subtracted by the $\Gt$ term. 

On the other hand, for $x > y$, we pick up the residue at the double pole $\kminus_{b}$ which requires us to differentiate. Using the product rule, we pick up the other half of the leading order result which is then subtracted by $\Gt$ as well as new term
\begin{equation}
- \int \frac{dx d\kperp}{2(2\pi)^3 x (1-x)} \; \frac{dy d\pperp}{2 (2\pi)^3 y (1-y)} \psi^*(x,\kperp) \theta (x - y) \frac{1}{2 R^+}\frac{\partial}{\partial \kminus_{b}} \tilde{V}(\kminus_{b}, x, \kperp; y, \pperp) \psi(y,\pperp).
\end{equation}
The relevant piece of the second term in Eq. \eqref{deltaN}, is evaluated identically up to $\{k \leftrightarrow p\}$. Thus we produce instead a factor of 
\begin{equation}
\theta(y-x) \frac{\partial}{\partial \pminus_{b}} \tilde{V}(\pminus_{b}, y, \pperp; x, \kperp). \notag
\end{equation} 
To make the end result symmetrical, we merely decode the notation: 
$\tilde{V}(\pminus_{b}, y, \pperp; x, \kperp) = \tilde{V}(\kminus_{a}, x, \kperp; y, \pperp)$. Now combining the two terms and their relevant 
$\theta$ functions, we can rewrite the result using the explicit form of the one-boson exchange potential, namely
\begin{equation} \label{nonN}
 \delta N = \int \frac{dx d\kperp}{2(2\pi)^3 x (1-x)} \; \frac{dy d\pperp}{2 (2\pi)^3 y (1-y)} \psi^*(x,\kperp) \Bigg( - \frac{\partial}{\partial M^2}      V(x, \kperp; y, \pperp) \Bigg)  \psi(y, \pperp).  
 \end{equation} 
 Thus although the covariant derivative's action on the potential vanishes, we can manipulate the correction to the normalization into the form of a derivative's action on the light front, time ordered potential. With this form, we can compare to the familiar nonvalence probability to be discussed in Appendix C (in the frame where $R^+ = R^- = M/\sqrt{2}$ and $M/\sqrt{2}$ is the eigenvalue of the light front Hamiltonian denoted $\pminus$ below).

 \section*{Appendix B: Instantaneous kernel} 
 We have seen in section \ref{current}, that crossing the interaction to interpret the non-wave function vertices which occur in evaluating form factors leads to contributions which are subtracted at the next order in perturbation theory. In our model, we have seen that the correct diagrams in the reduction scheme are light front time ordered and there are no extra entangled contributions to deal with from hadronization off quark lines. This still leaves a provocative question. What about the case of QCD${}_{1+1}$ considered in \cite{Einhorn:1976uz} where such interaction crossing symmetry is used?  

 The effective two-body interaction in the 't Hooft model is instantaneous in light cone time, and hence is independent of the minus momenta. Here, we specialize to the case of an instantaneous light front potential (or equivalently the leading piece of an instantaneous light front reduction scheme\footnote{Additionally it is true in models where the vertex is described by a contact interaction. One can probe the nonvalence sector although there are no true 
higher Fock space components.}) to clear up the apparent disparity in dealing with the non-wave function vertices. 

From our comments in the text (section \ref{intuitive}), we see that since our effective interaction is instantaneous, the fully covariant Bethe-Salpeter vertex function $\Gamma$ is light-front energy independent. In light of this observation, we can say a number of things: e.g. the Bethe-Salpeter wave function $\Psi$ only has "valence" poles which stem from the propagator; thus, in particular, no higher Fock space components can be generated by carrying out minus momentum integrals in expressions for form factors. This is part of the clue for dealing with nonvalence contributions directly in this specialized situation. We shall elaborate this for the case of form factors. Our notation is that set out in considering leading order expressions for form factors in section \ref{current}, although we mainly write down only schematic equations in this Appendix. We omit total momentum labels where there is no ambiguity, 
i.e.~when they are all the same.

 With no minus momentum dependence in the vertex function, we can write the Bethe-Salpeter equation \eqref{BS} immediately in a form similar to Eq. \eqref{gammared} by performing the $\kminus$ integral which picks up contributions identical to Eq. \eqref{g}
 \begin{equation} \label{mini}
 |\gamma \rangle = v g | \gamma \rangle,
 \end{equation}
 where $v$ is to represent the instantaneous interaction, which by nature is already three dimensional. We are able to make an identical statement 
 (to one made earlier) about nonvalence contributions to the vertex, i.e. since $g\propto\theta[x(1-x)]$, $\gamma$ for $0<x<1$ uniquely determines 
 $\gamma$ everywhere. 

 We can define a wave function in terms of the reduced vertex in the usual way, $|\psi\rangle = g |\gamma\rangle$. The reduction in \eqref{mini} naturally shows this wave function is indeed the $x^+=0$ projection of the Bethe-Salpeter equation since
 \begin{align} 
 |\psi\rangle & = \Big| \; | \Psi \rangle = \Big| G | \Gamma \rangle \notag \\
 		 & = \Big| G \Big| \; |\gamma\rangle = g |\gamma\rangle.
 \end{align}
 In the second line, we used the fact that the reduced vertex has no minus momentum dependence and is thus already three dimensional. In this form it is easy to see that iteration of the Bethe-Salpeter equation (like that carried out in section \ref{intuitive}) leads to nothing new; after all, there are no poles of the vertex function to better approximate. 

 Now consider the electromagnetic form factor for this special case. It is defined in terms of a matrix element between bound states of the irreducible five-point function
 \begin{equation}
 \langle \Psi_{f} | \Gamma^\mu | \Psi_{i} \rangle.
 \end{equation}
 As our concern is only with the non-wave function vertex, we can ignore the gauged instantaneous interaction's contribution to $\Gamma^\mu$. Similarly, we need only consider the bare term in the electromagnetic three-point function to be confronted with such a nonvalence contribution. 
 The relevant contribution under consideration from the above equation is then
 \begin{equation} \label{scheme}
 \langle\gamma_{f}| \; \Big| G(P_f) d^{+}d_{2}^{-1} G(P_i) \Big| \; |\gamma_{i}\rangle, 
 \end{equation}
 having used $|\Psi\rangle = G |\Gamma\rangle = G \big| \;|\gamma\rangle$. This form is identical to the leading order expression considered in section \ref{current} for the form factor. Having already evaluated this expression, we know there are two contributions, that of Figure \ref{ftri} for $x>\zeta$ and the nonvalence piece in Figure \ref{fZ}. We are then in the position to iterate Eq. \eqref{mini} to extend the non-wave function vertex's definition by crossing. 
 Above, however, we learned to treat such a procedure with care since in the process an additional factor of the presumably weak coupling is added. 

 To convincingly derive the form of the non-wave function vertex in this instantaneous model, we first iterate the Bethe-Salpeter equation for the final state (any iteration for the initial state will only get reduced back into the wave function). Using
 \begin{equation}
 \langle \Psi_{f} | = \langle \Gamma_{f} | G(P_f) = \langle \Gamma_{f} | G(P_f) V(P_f) G(P_f),  
 \end{equation}
 we can then write the current matrix element as
 \begin{equation}
 \langle \psi_{f} | g^{-1}(P_f) \Big| G(P_f) V(P_f) G(P_f) d^+ d_{2}^{-1} G(P_i) \Big| \; |\gamma_{i}\rangle.
 \end{equation}
 Now we use the fact that our interaction is instantaneous to trivially do the minus momentum loop integral added from the iteration of the Bethe-Salpeter 
 equation. The above equation converts to
 \begin{equation} \label{aha}
 \langle\psi_{f}|\; v(P_f) \; \Big| G(P_f)  d^+ d_{2}^{-1} G(P_i) \Big| \; |\gamma_{i}\rangle.
 \end{equation}
The final minus momentum integration produces two contributions. Firstly for $\zeta< x< 1$, we are able to use the bound state equation for the final state wave function since $0<y<1$. Inserting $\langle\psi_{f}|v = \langle\gamma_{f}|$, we arrive back at the original expression in this region Eq. \eqref{scheme}. 
On the other hand, when $0<x<\zeta$, we have $y<0$ and appealing to the bound state equation is of no avail. However, what remains in equation \eqref{aha} is precisely what we would have obtained by extending the definition of the vertex via \eqref{mini}. 

Therefore, in the case of an instantaneous interaction, the non-wave function vertices can be interpreted by crossing symmetry as is done in \cite{Einhorn:1976uz}. In such a situation, there are no higher Fock space contributions contained in the covariant Bethe-Salpeter wave function. One would also have
non-vanishing, non-wave function vertices to leading order in an instantaneous reduction scheme of the Bethe-Salpeter equation (i.e. using another form of  $\Gt$ than considered in section \ref{formal} to do the instantaneous reduction). Higher order terms, however, would consist of entangled instantaneous terms and higher Fock components. 
From our analysis above, we know the instantaneous pieces would cancel in favor of higher Fock components---making 
their pressence at leading order superficial. 
 Additionally the instantaneous reduction of the kernel would not have the desirable feature unique to light front time ordered perturbation theory---that is, at higher orders one systematically adds more particles propagating at a given instant (in light cone time) which has been demonstrated to lead to quicker convergence \cite{Sales:1999ec,Schoonderwoerd:1998pk} even over (instant) time ordered perturbation theory. Finally we remark 
that although certainly not ideal, an instantaneous interaction approximation could give some new information about the nonvalence 
structure of timelike form factors and timelike decays in the light front formalism.

\section*{Appendix C: Two-body projection operators}
The results of this paper can similarly be achieved directly from ``old fashioned'' time ordered perturbation theory in a form which utilizes
projecting onto the two-body subspace of the full Fock space. As the light front Hamiltonian generates light cone time translation, not surprisingly time ordered perturbation theory stems from the Hamiltonian.
For a nice, complete discussion of this formalism for our scalar model, see \cite{Cooke:2000ef}. 
In this Appendix, we show how to derive higher Fock space components in this 
formalism, thereby demonstrating the generation of higher components from the lowest sector we found indirectly for GPD's and form factors in
section \ref{gpd}. 

We write the Hamiltonian as a sum of a free piece and an interacting piece which carries an explicit power of the weak coupling $g$. In an obvious notation 
this is
\begin{equation}
P^- = P^-_o + g P_{I}^-.
\end{equation}
The free term $P^-_o$ is diagonal in the Fock state basis, while the interaction generally mixes components of different particle number (in the scalar model we consider above, the interaction is completely off-diagonal since there are no instantaneous terms). Let us suppose that in the full Fock basis, we have an eigenstate of the light-front Hamiltonian, i.e.
\begin{equation}
\Big( P^-_o + g P_{I}^- \Big) \psiket = \pminus \psiket, 
\end{equation}
where the eigenvalue is labeled by $\pminus$. Since the coupling is presumed small, the mixing of Fock components with a large number of 
particles will be small. Thus one imagines our bound state will be dominated by the two-body Fock component. 

To make this observation formal, 
we define projections operators on the Fock space $\pe$ and $\qu$ in the usual sense (i.e. $\pe^2 = \pe$, $\qu^2 = \qu$, $\pe + \qu =1$, \emph{etc}.). 
The operator $\pe$ projects out only the two-particle subspace of the full Fock space and hence $\qu$ projects out the compliment. Let us define the action
of these operators on our eigenstate
\begin{align}
\pe \psiket & = \psitwoket \\
\qu \psiket & = \psiquket.
\end{align}
Given $\pe + \qu = 1$, we have immediately that $\psiket = \psitwoket + \psiquket$. We now wish to derive an equation for the two-body state
in terms of effective two-body operators. First we note
\begin{equation} \label{start}
\pe P^- \psiket = \pminus \pe \psiket = \pminus \psitwoket,
\end{equation}
and then write $\pe P^- 1 = \pe P^- (\pe + \qu) = P^-_{\pe \pe} \pe + P^-_{\pe \qu} \qu$, where we have defined the following notation
for any operators $A$ and $B$, $P^-_{A B} \equiv A P^- B$. Using this operator relation in Eq. \eqref{start} yields
\begin{equation} \label{one}
P^-_{\pe \qu} \psiquket = \Big( \pminus - P^-_{\pe \pe}  \Big) \psitwoket.
\end{equation} 
Now we use the same procedure on the state $\qu P^- \psiket$ to find
\begin{equation} \label{two}
P^-_{\qu \pe} \psitwoket = \Big( \pminus - P^-_{\qu \qu}  \Big) \psiquket.
\end{equation}

Combining Eqs. \eqref{one} and \eqref{two}, we arrive at the following equation for the two-body Fock component
\begin{equation}
P^-_{\text{eff}} \psitwoket = \pminus \psitwoket,
\end{equation}
where the effective two-body Hamiltonian is
\begin{align} \label{veff}
P^-_{\text{eff}} & \equiv P^-_{\pe \pe} + V_{\text{eff}} \\
	& = P^-_{\pe \pe} + P^-_{\pe \qu} \frac{1}{\pminus - P^-_{\qu \qu}} P^-_{\qu \pe}.
\end{align}
The effective two-body interaction $V_{\text{eff}}$ defined in equation \eqref{veff} is dependent upon the energy $\pminus$ since we have 
suppressed the degrees of freedom of the $\qu$ subspace. This yields a relation between the $\qu$-space probability (i.e. the nonvalence contribution)
and the effective interaction
\begin{equation} \label{quspace}
\langle \psi_{\qu} \psiquket = - \frac{\partial}{\partial \pminus} \psitwobra V_{\text{eff}} \psitwoket.
\end{equation}

In a weak binding limit, we can series expand the effective interaction in powers of the coupling and thereby re-derive the results of section \ref{intuitive} for the light front reduced potential. Given that every boson emitted must be absorbed in the two quark sector, we can have only an even number of interactions and hence
\begin{equation}
V_{\text{eff}} = \pe g P^-_{I} \qu \frac{1}{\pminus - P_{o}^-} \sum_{n=0}^{\infty} \Bigg( \frac{g P^-_{I}}{\pminus - P^-_{o}} \Bigg)^{2 n} \qu g P^-_{I} \pe.
\end{equation}
So, for example, at leading order we have all possible ways to propagate from the two-body sector and back with only two interactions in between. The diagrams in Figure \ref{fOBEP} correspond to the two possibilities distinguished by the action of $\frac{1}{\pminus - P^-_{o}}$ between interactions. At the next order, we have all possible ways to propagate from two bodies to two bodies with four interactions in between (see Figure \ref{fv2}), \emph{etc}.

 To generate higher Fock components from the two-body sector, we necessarily must look at the $\qu$ space state,
 \begin{equation}
 \psiquket = \frac{1}{\pminus - P^-_{\qu \qu}} P^-_{\qu \pe} \psitwoket. 
 \end{equation}   
 To generate an $n$-body Fock component from this state, we merely act with an $n$-body projection operator which we shall denote $\qu_{n}$. Similar to the above, we expand in powers of the coupling to find
\begin{equation}
| \psi_{n} \rangle = \qu_{n} \frac{1}{\pminus - P^-_{o}} \sum_{n = 0}^{\infty} \Bigg(  \frac{g P^-_{I}}{\pminus - P^-_{o}} \Bigg)^n \qu g P^-_{I} \psitwoket.
\end{equation}
For example, the leading order three-body state is obtained by attaching a boson to a quark line in the only two possible ways (and adding the light front energy denominator at the end). With these three body states, we can consider all possible three-to-three overlaps that would contribute to the form factor. These
are depicted in Figure \ref{ftri2} (with the exception of a quark 
self-interaction).  The four-body sector is richer since there are two boson, two quark states as well as four quark states. The two-to-four overlaps required for GPD's must have four quarks. At leading order, we generate the diagrams encountered above in Figure \ref{fZZZ}. Not surprisingly directly applying 
time ordered perturbation theory from a light front Hamiltonian agrees with our derivation above from covariant perturbation theory in the Bethe-Salpeter formalism.

\end{document}